\documentclass{Buried-layer-template}
\usepackage[super,comma,sort&compress]{natbib}
\usepackage[T1]{fontenc}
\usepackage[flushleft]{threeparttable}
\usepackage{hyperref}
\hypersetup{colorlinks,citecolor=blue}
\usepackage{mathptmx}
\usepackage{multirow}
\usepackage{amsmath}

\newcommand{\sous}[1]{\ensuremath{_{\textrm{#1}}}}

\begin{document}
\setlength{\parskip}{1em}
\pagestyle{plain}
\justifying

\title{Thermal conductivity measurements of sub-surface buried substrates by steady-state thermoreflectance}

\maketitle


\author{Md Shafkat Bin Hoque, Yee Rui Koh, Kiumars Aryana, Eric Hoglund, Jeffrey L. Braun, David H. Olson, John T. Gaskins, Habib Ahmad, Mirza Mohammad Mahbube Elahi, Jennifer K. Hite, Zayd C. Leseman, W. Alan Doolittle, and Patrick E. Hopkins}


\dedication{}

\begin{affiliations}
Md Shafkat Bin Hoque, Yee Rui Koh, Kiumars Aryana, Jeffrey L. Braun, David H. Olson, John T. Gaskins\\
Department of Mechanical and Aerospace Engineering, University of Virginia, Charlottesville, Virginia 22904, USA\hfill

Eric Hoglund\\
Department of Materials Science and Engineering, University of Virginia, Charlottesville, Virginia 22904, USA\hfill

Jennifer K. Hite\\
U.S. Naval Research Laboratory, Washington, D.C. 20375, U.S.A.

Mirza Mohammad Mahbube Elahi\\
Department of Electrical and Computer Engineering,  University of New Mexico, Albuquerque, New Mexico 87131, USA\hfill

Zayd C. Leseman\\
Department of Mechanical Engineering,  King Fahd University of Petroleum $\&$ Minerals, Dhahran, Eastern Province 31261, Saudi Arabia\hfill

Habib Ahmad, W. Alan Doolittle\\
School of Electrical and Computer Engineering, Georgia Institute of Technology, Atlanta, GA, 30332, United States\hfill

Patrick E. Hopkins\\
Department of Mechanical and Aerospace Engineering, University of Virginia, Charlottesville, Virginia 22904, USA\\
Department of Materials Science and Engineering, University of Virginia, Charlottesville, Virginia 22904, USA\\
Department of Physics, University of Virginia, Charlottesville, Virginia 22904, USA\\
Email: phopkins@virginia.edu\hfill

\end{affiliations}

\newpage
\textbf{\Large Abstract}
\begin{abstract}\\
Measuring the thermal conductivity of sub-surface buried substrates are of significant practical interests. However, this remains challenging with traditional pump-probe spectroscopies due to their limited thermal penetration depths (TPD). Here, we experimentally and numerically investigate the TPD of recently developed optical pump-probe technique steady-state thermoreflectance (SSTR) and explore its capability for measuring the thermal properties of buried substrates. The conventional definition of the TPD does not truly represent the upper limit of how far beneath the surface SSTR can probe. For estimating the uncertainty of SSTR measurements of a buried substrate a priori, sensitivity calculations provide the best means. Thus, detailed sensitivity calculations are provided to guide future measurements. Due to the steady-state nature of SSTR, it can measure the thermal conductivity of buried substrates typically inaccessible by traditional pump-probe techniques, exemplified by measuring three control samples. We also discuss the required criteria for SSTR to isolate the thermal properties of a buried film. Our study establishes SSTR as a suitable technique for thermal characterizations of sub-surface buried substrates in typical device geometries.
\end{abstract}



\section{Introduction}
Thin films with thicknesses ranging from nanometer to micrometer length scales have become an integral part of transistors,\cite{park2008thin,tokunaga2010thin} thermoelectrics,\cite{venkatasubramanian2001thin} optical coatings,\cite{xi2007optical} solar cells,\cite{peumans2003small} and memory devices.\cite{yang2004organic,sung2015systematic} As the device efficiency and reliability are often dictated by the thermal performance, it is of crucial importance to properly characterize the thermal properties of the thin films and substrates.\cite{dames2013measuring} Traditional non-contact, optical pump-probe techniques such as time-domain thermoreflectance (TDTR)\cite{cahill2004analysis,schmidt2008pulse,feser2014pump,jiang2018tutorial} and frequency-domain thermoreflectance (FDTR)\cite{schmidt2009frequency} are widely used to measure the thermal conductivity of thin films.\cite{zhao2016measurement} However, TDTR and FDTR often cannot measure the thermal conductivity of buried substrates located beyond 1 $\mu$m due to shorter thermal penetration depths (usually < 1 $\mu$m). \cite{zhu2010ultrafast,braun2019steady} The recently developed optical pump-probe technique steady-state thermoreflectance (SSTR) offers a solution to this issue as its thermal penetration depth can be much larger than those produced during TDTR and FDTR measurements.\cite{braun2019steady,koh2020bulk} Therefore, a detailed study into the thermal penetration depth of SSTR technique and its ability to measure the thermal conductivity of sub-surface buried substrates is highly warranted.

Using the principle of thermoreflectance,\cite{rosei1972thermomodulation} SSTR employs co-axially focused pump and probe beams from continuous wave (CW) lasers to directly measure the thermal conductivity of a material by applying Fourier's law. Schematics of the SSTR measurement configuration and principle are shown in Figure 1(a) and 1(b), respectively. Using a low modulation frequency (\textit{f}), the pump laser generates a periodic heat flux at the sample surface for an extended period. The low modulation frequency provides enough time for the system to reach steady-state. The probe beam then measures the resultant steady-state temperature rise by monitoring the reflectance change with a balanced photodetector and a lock-in amplifier. A linear relation between the heat flux and temperature rise is established by varying the pump power and monitoring the reflectance change at each pump power. From this relation, the thermal conductivity of any material can be determined by using Fourier's law. The thermal conductivity tensor measured by SSTR is different from that of TDTR or FDTR.
For bulk materials, whereas TDTR and FDTR usually measure the cross-plane thermal conductivity, SSTR measures $\sqrt{k_rk_z}$, where, \textit{k$_z$} and \textit{k$_r$} are cross-plane and in-plane thermal conductivity, respectively.\cite{braun2019steady} 

\begin{figure}
\centering
\includegraphics[scale=0.54]{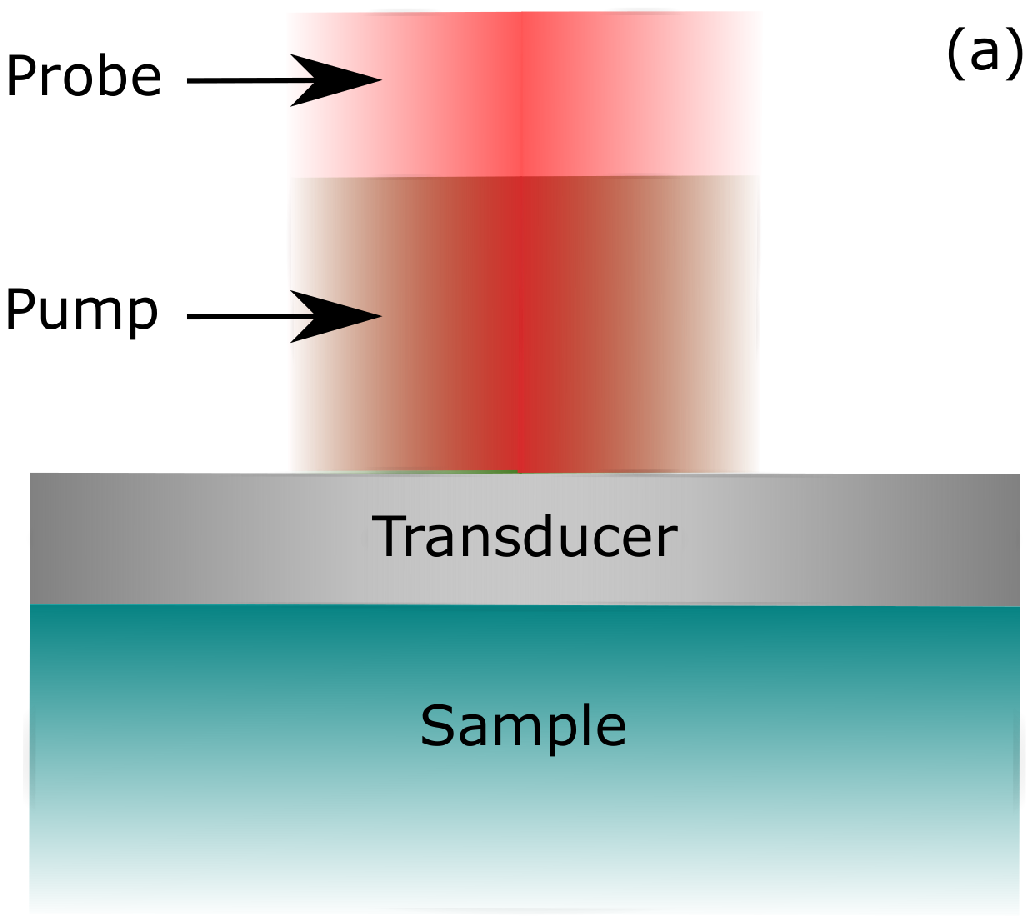}\hfill
\includegraphics[scale=0.35]{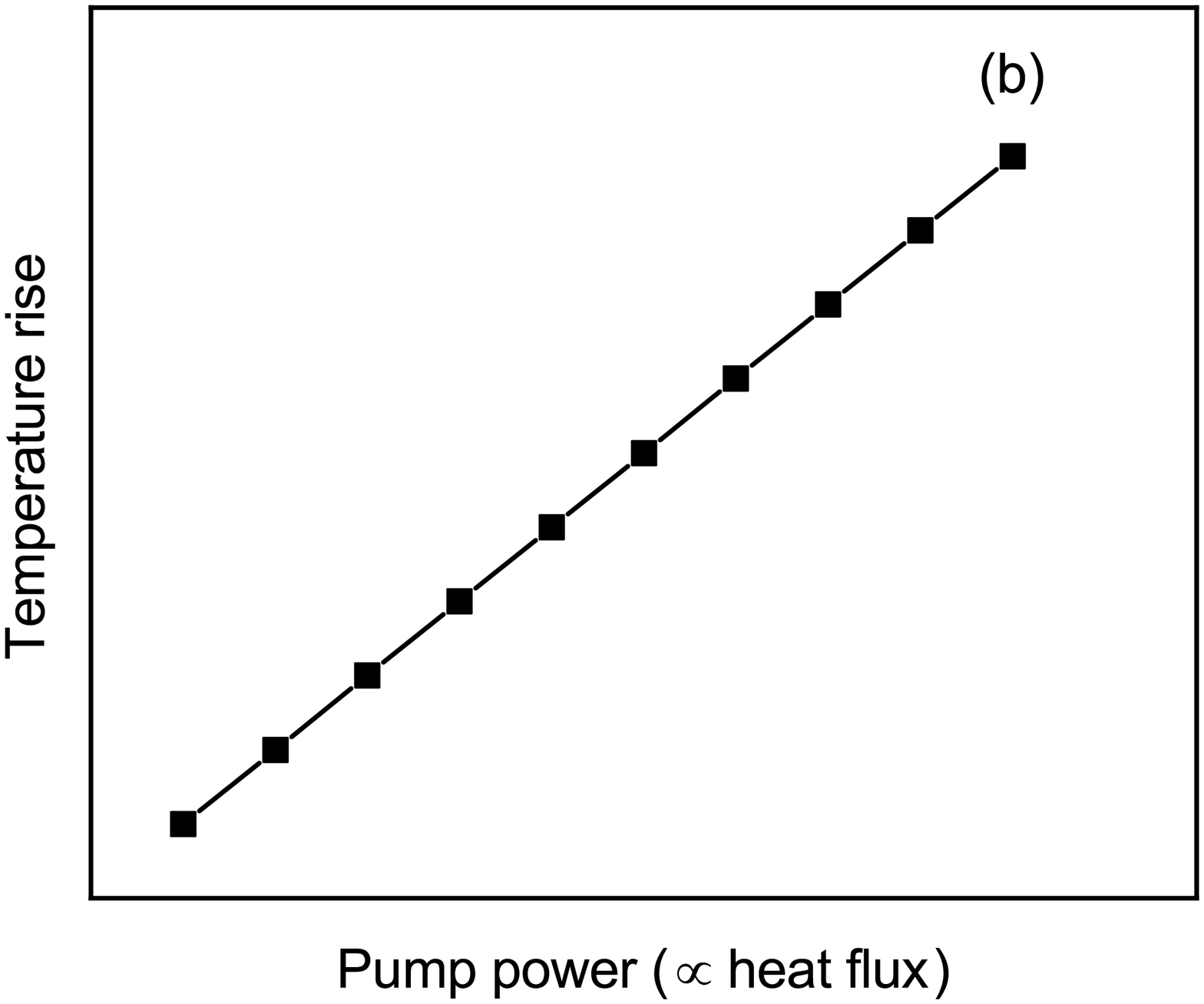}\hfill
\caption{Schematics of the SSTR measurement (a) configuration and (b) principle.}
\label{fig:1}
\end{figure}

In alignment with TDTR and FDTR, in SSTR, the thermal penetration depth (TPD) is defined as the distance normal to the surface at which the temperature drops to the 1/e value of the maximum surface temperature (T\sous{max}). \cite{koh2007frequency, braun2017upper,braun2018steady,olson2019spatially} According to this definition, the 1/e$^2$ heater (pump) radius represents the upper limit of the TPD when the modulation frequency is low (\textit{f} $\rightarrow$ 0), such as in SSTR.\cite{olson2019spatially} However, such a description of the SSTR TPD fails for multilayer material systems (i.e., a thin film on a substrate). In such systems, the TPD can change widely based on the ratio of thin film to substrate thermal conductivity, and the thermal boundary conductance (\textit{G}) between the thin film and substrate. This is further complicated by the presence of thin metal film transducers at the sample surface, which are often a requirement in optical pump-probe techniques for optothermal transduction.\cite{wilson2012thermoreflectance,wang2010thermoreflectance,rost2017hafnium,wang2016thermal,radue2018hot} 

In this study, we numerically and experimentally analyze the TPD definition of SSTR measurements. We also discuss the implications of metal film transducers, thermal boundary conductances, and role of multilayer material systems on the heuristic approximations for the SSTR TPD. Notably, our experimental results indicate that although the traditional TPD definition provides a convenient estimate of how far beneath the surface SSTR can probe, it does not represent the absolute upper limit of SSTR probing depth. In specific cases, SSTR can probe beyond both the 1/e temperature drop distance and the heater radius. Furthermore, the thermal conductivity measurements of buried layers or substrates by SSTR are not solely dictated by the TPD. The uncertainty associated with such measurements are often governed by the transducer thermal conductivity, the thermal boundary conductances, and the thermal resistances offered by different layers of the multilayer material system. Thus when determining whether the material of interest at some depth under the surface is measurable within acceptable limits of uncertainty, sensitivity calculations provide the best means. Moreover, we show that due to the continuous wave nature of the pump laser source and low modulation frequency, SSTR can measure the thermal conductivities of buried substrates that are typically inaccessible by TDTR and FDTR. This is illustrated by measuring the thermal conductivities of buried substrates in three different samples: i) $\sim$130 nm amorphous silicon dioxide (a-SiO$_{2}$) thin film on a silicon (Si) substrate, ii) $\sim$2.05 $\mu$m gallium nitride (GaN) thin film on a n-type GaN substrate, and iii) $\sim$2 $\mu$m aluminum nitride (AlN) thin film on a sapphire substrate. In addition, it is also established that using large 1/e$^2$ pump and probe radii, SSTR can measure the thermal conductivity of highly resistive buried films.

\section{Results and discussion}
\subsection{SSTR thermal penetration depth of a 2-layer system}

We first review how the TPD of SSTR changes as a function of substrate thermal conductivity in a 2-layer system: metal transducer/substrate. The substrate here represents a bulk isotropic material. The TPD is calculated by solving the cylindrical heat diffusion equation, detailed descriptions of which are given elsewhere.\cite{braun2017upper} For these calculations, 1/e$^2$ pump and probe radii (\textit{r}$_{o}$ and \textit{r}$_{1}$, respectively) of 10 $\mu$m are used. The modulation frequency is chosen to be 100 Hz as it represents a realistic value usable in an experiment. We further assume that all the energy is absorbed in an infinitesimal thin layer on the surface (i.e., surface boundary condition).

\begin{figure}[hbt!]
\centering
\includegraphics[scale=0.35]{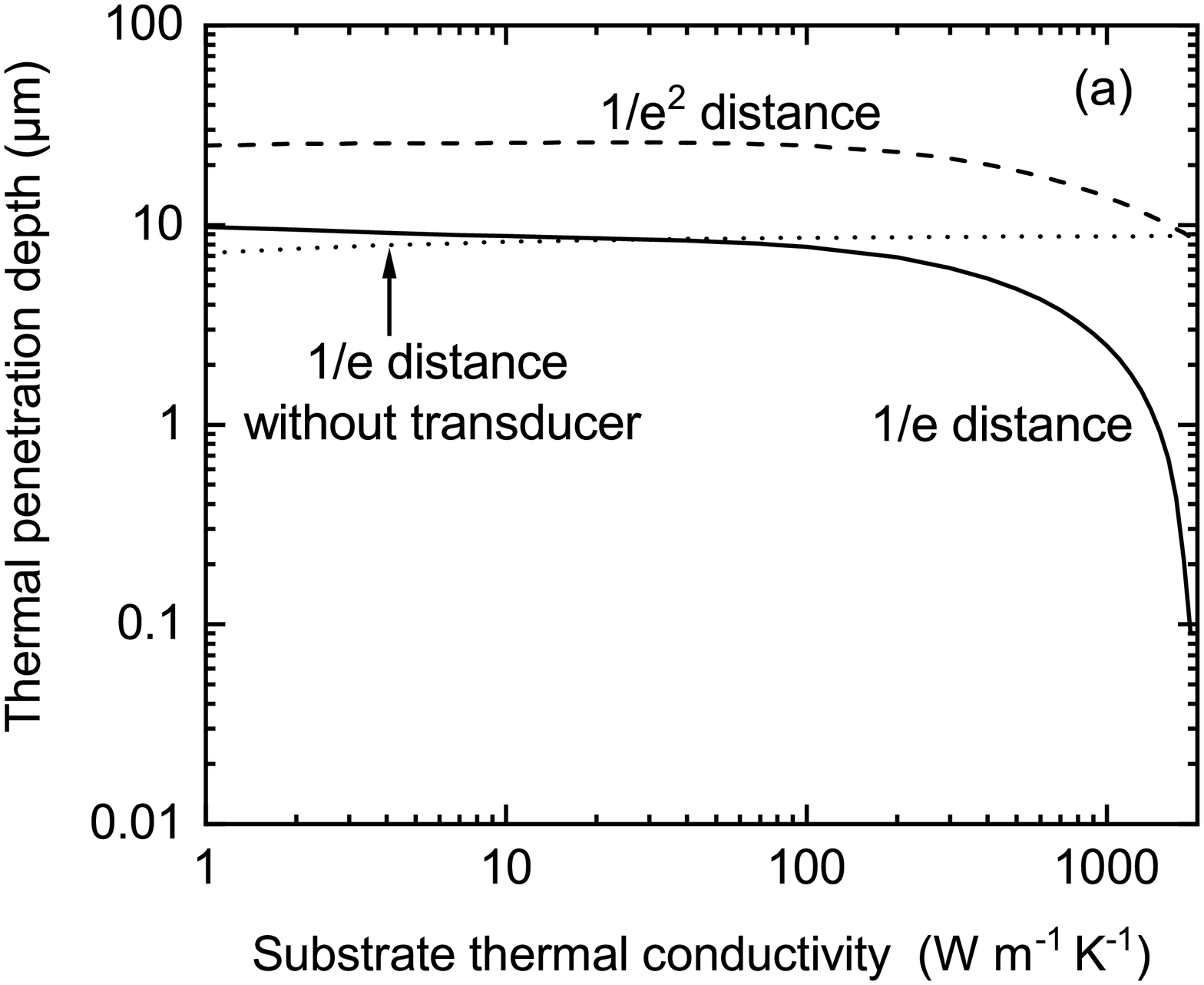}\hfill
\includegraphics[scale=0.35]{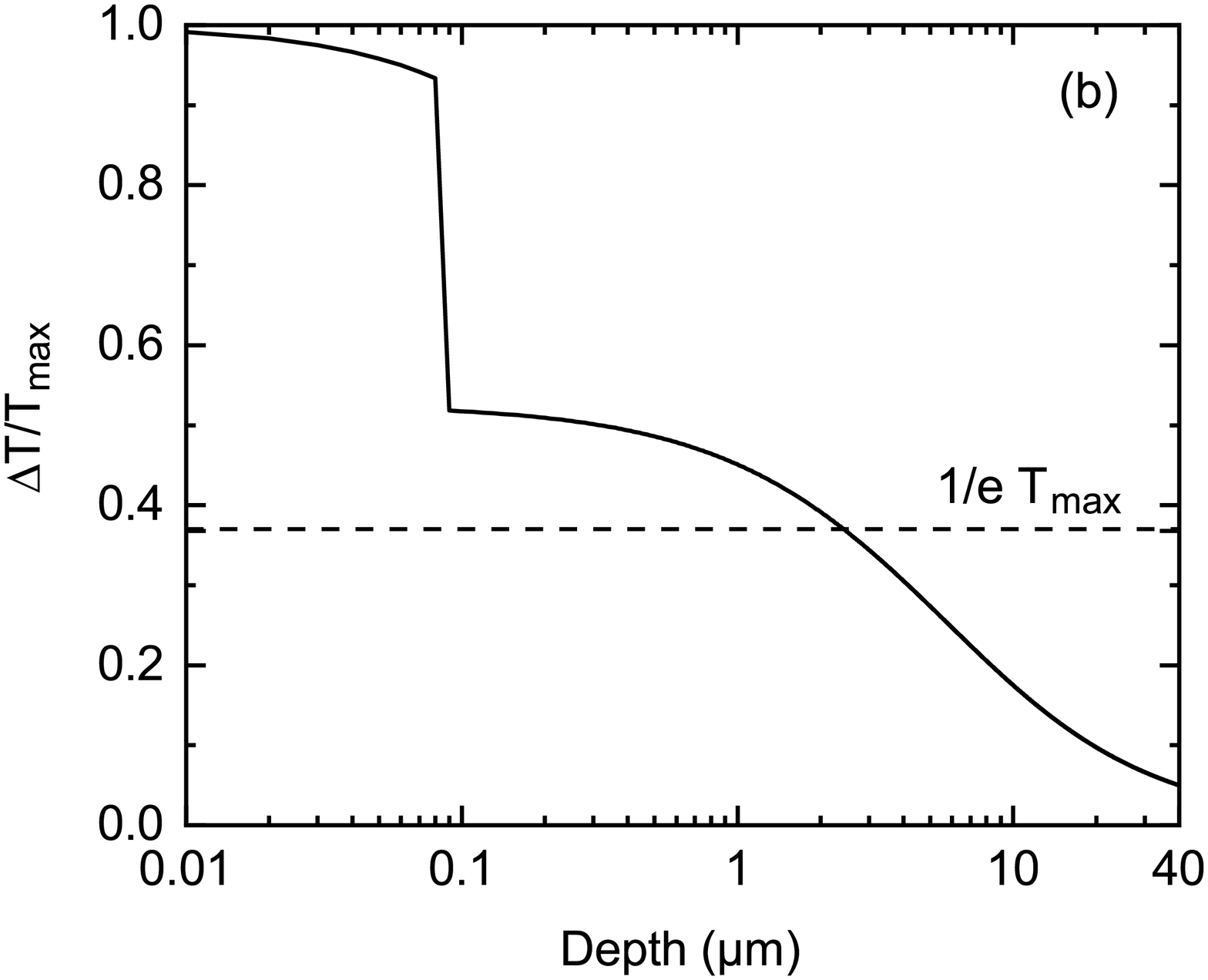}\hfill
\caption{(a) Thermal penetration depth as a function of substrate thermal conductivity for a 2-layer system: metal transducer/substrate. (b) Normalized temperature drop ($\triangle$T/T\sous{max}) as a function of depth for a substrate thermal conductivity of 1000 W m$^{-1}$ K$^{-1}$. The calculations correspond to \textit{f} = 100 Hz, \textit{d}$_1$ = 80 nm, \textit{r}$_0$ = \textit{r}$_1$ = 10 $\mu$m, \textit{k}$_{1}$ = 100 W m$^{-1}$ K$^{-1}$, \textit{C}$_{V, 1}$ = \textit{C}$_{V, 2}$ = 2 MJ m$^{-3}$ K$^{-1}$, and \textit{G}$_{1}$ = 200 MW m$^{-2}$ K$^{-1}$. Here, \textit{d} and \textit{C}$_{V}$ represent thickness and volumetric heat capacity, respectively.}
\label{fig:2}
\end{figure}

In Figure 2(a), the TPD corresponding to the 1/e temperature drop distance from the surface is presented for two scenarios, with and without the inclusion of a transducer. When no transducer is present, the change in the TPD is very small with respect to the substrate thermal conductivity. The small decrease in the TPD with substrate thermal conductivity reduction can be attributed to the choice of modulation frequency.  For a given pump and probe radii, the lower the substrate thermal conductivity, the longer it takes for the system to reach steady-state.\cite{braun2019steady} Thus, as the substrate thermal conductivity decreases, the system slightly deviates from the ideal steady-state condition (\textit{f} = 0). To keep the TPD constant, the modulation frequency needs to be lowered in accordance with the substrate thermal conductivity reduction. However, for the chosen modulation frequency of 100 Hz, the deviation from the ideal steady-state condition is quite small for the substrate thermal conductivities considered here and therefore, the system can still be reasonably approximated to be in steady-state.\cite{braun2019steady}

The presence of a transducer drastically changes the TPD. When the substrate thermal conductivity is low (< 10 W m$^{-1}$ K$^{-1}$), the TPD with the transducer is higher than the TPD without the transducer. This stems from the radial heat spreading in the transducer and a corresponding increase in the overall heater radius.\cite{olson2019spatially} On the other hand, when the substrate thermal conductivity is high (> 100 W m$^{-1}$ K$^{-1}$), the TPD with the transducer sharply deceases. The higher the substrate thermal conductivity is compared to that of the transducer, the lower the temperature rise is in the substrate for a given amount of heat flux. As a result, in high thermal conductivity substrates, a large temperature drop exists at the interface between the transducer and substrate. This is exemplified in Figure 2(b), where we present the normalized temperature drop as a function of depth for a substrate thermal conductivity of 1000 W m$^{-1}$ K$^{-1}$. In this example, the temperature decreases by nearly 41$\%$ at the transducer/substrate interface, leading to a TPD of 2.48 $\mu$m. 

We also calculate the distance normal to the surface at which the temperature drops to 1/e$^{2}$ value of maximum surface temperature and present it in Figure 2(a). It is evident that the 1/e$^{2}$ distance (calculated with a transducer) is much higher than the 1/e distance for all substrate thermal conductivities. This is to be expected as the temperature decay increases with depth.

\subsection{SSTR thermal penetration depth of a 3-layer system}

\begin{figure*}[hbt!]
\includegraphics[scale=0.325]{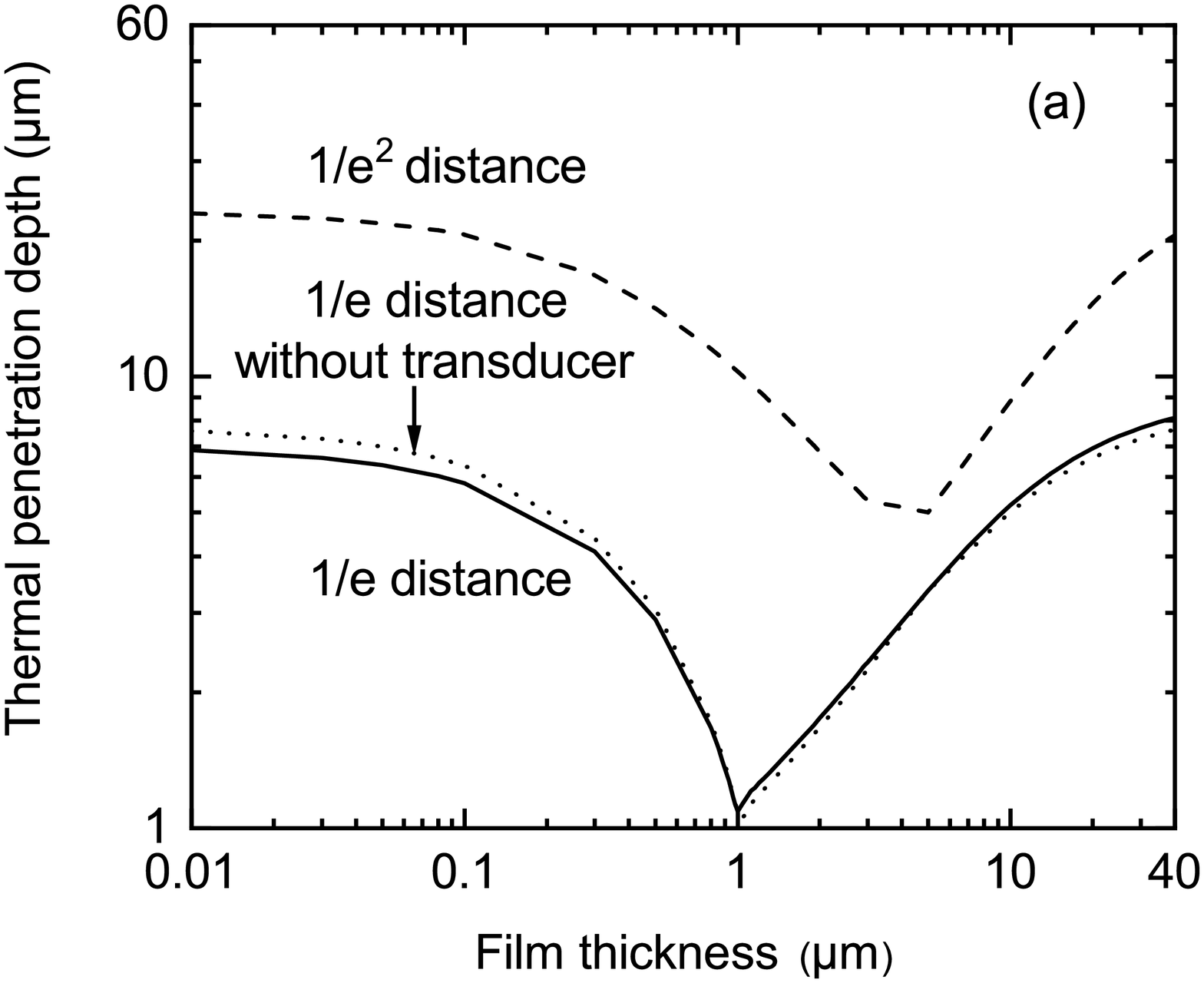}\hfill
\includegraphics[scale=0.325]{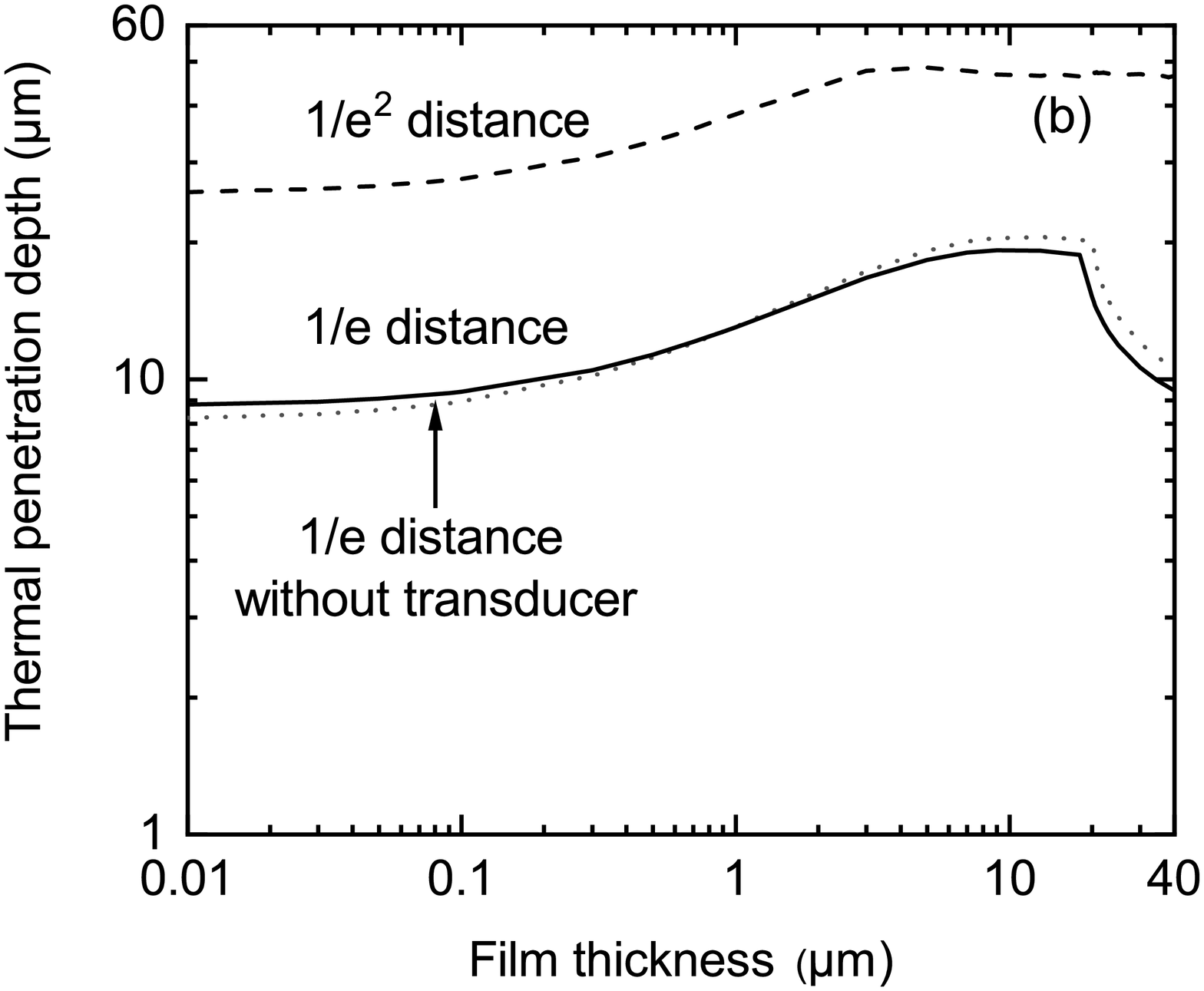}\hfill
\includegraphics[scale=0.32]{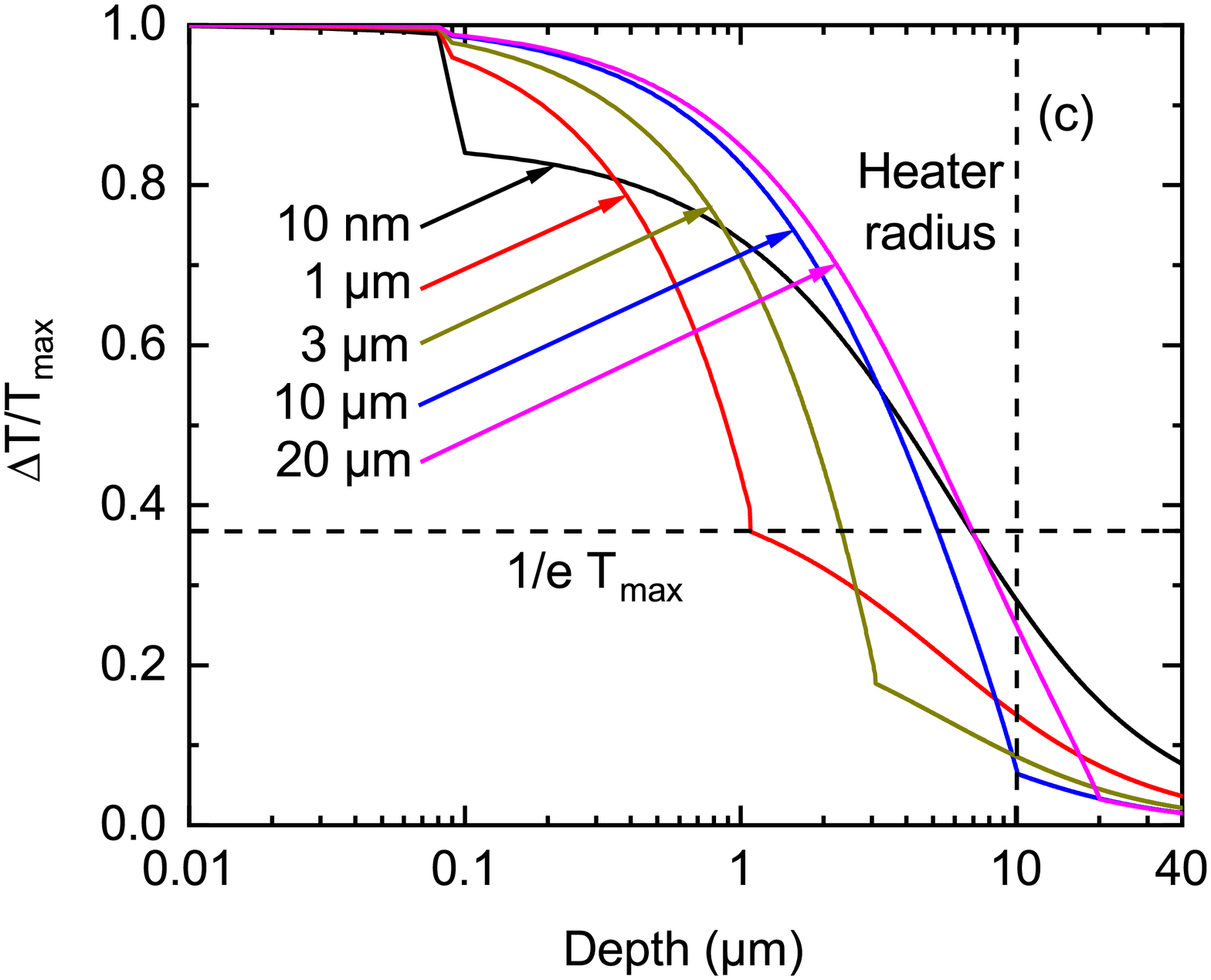}\hfill
\includegraphics[scale=0.32]{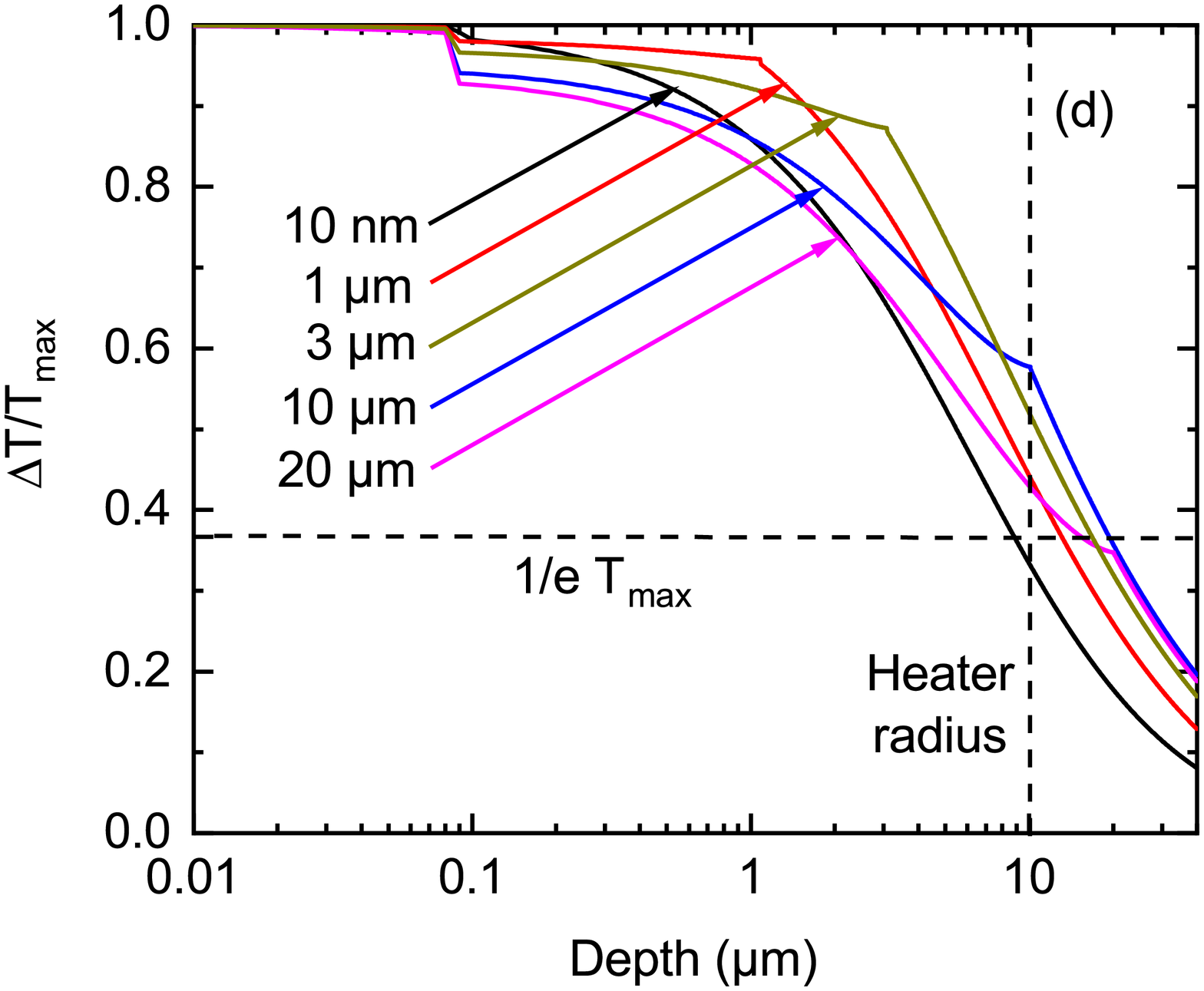}\hfill
\includegraphics[scale=0.29]{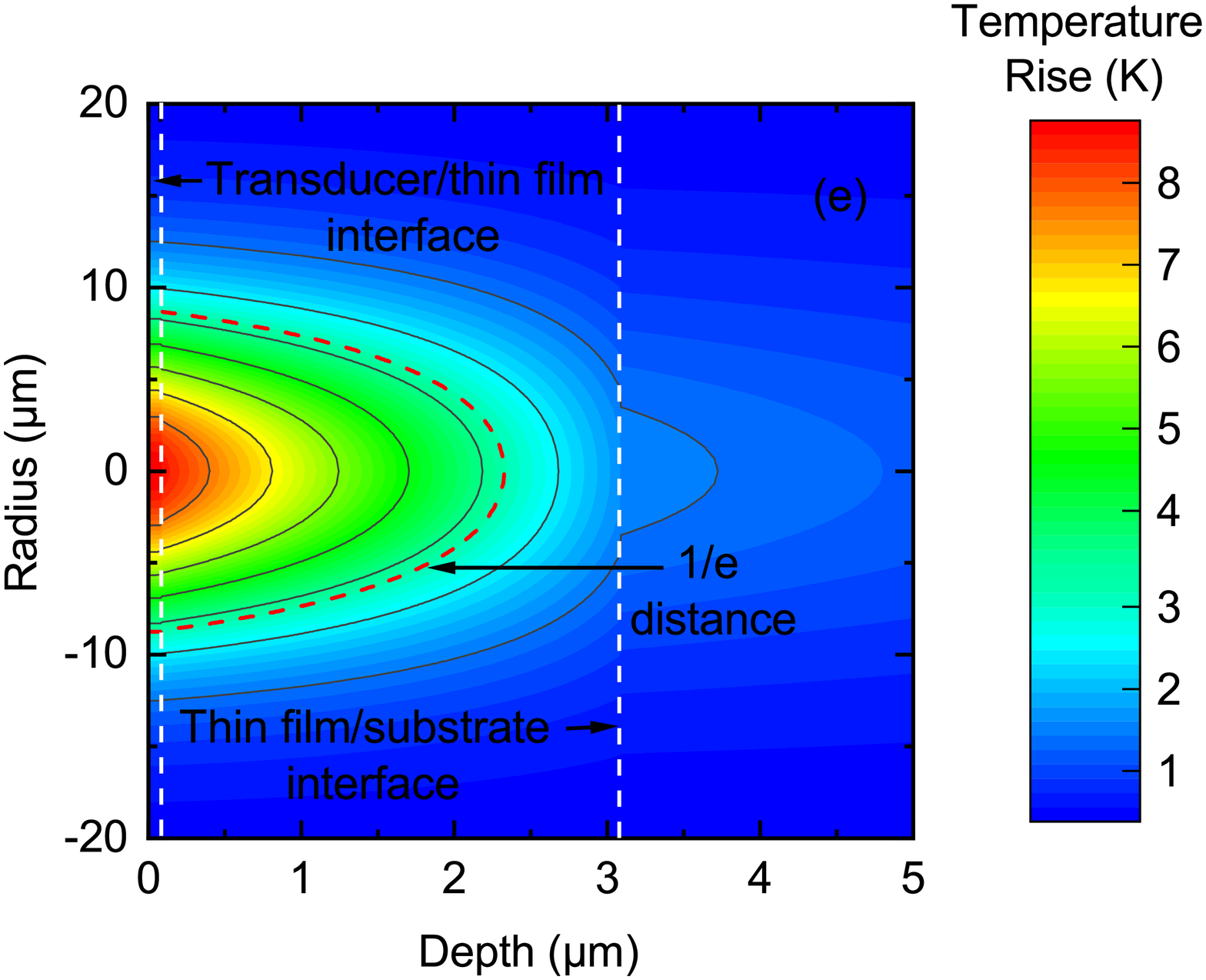}\hfill
\includegraphics[scale=0.29]{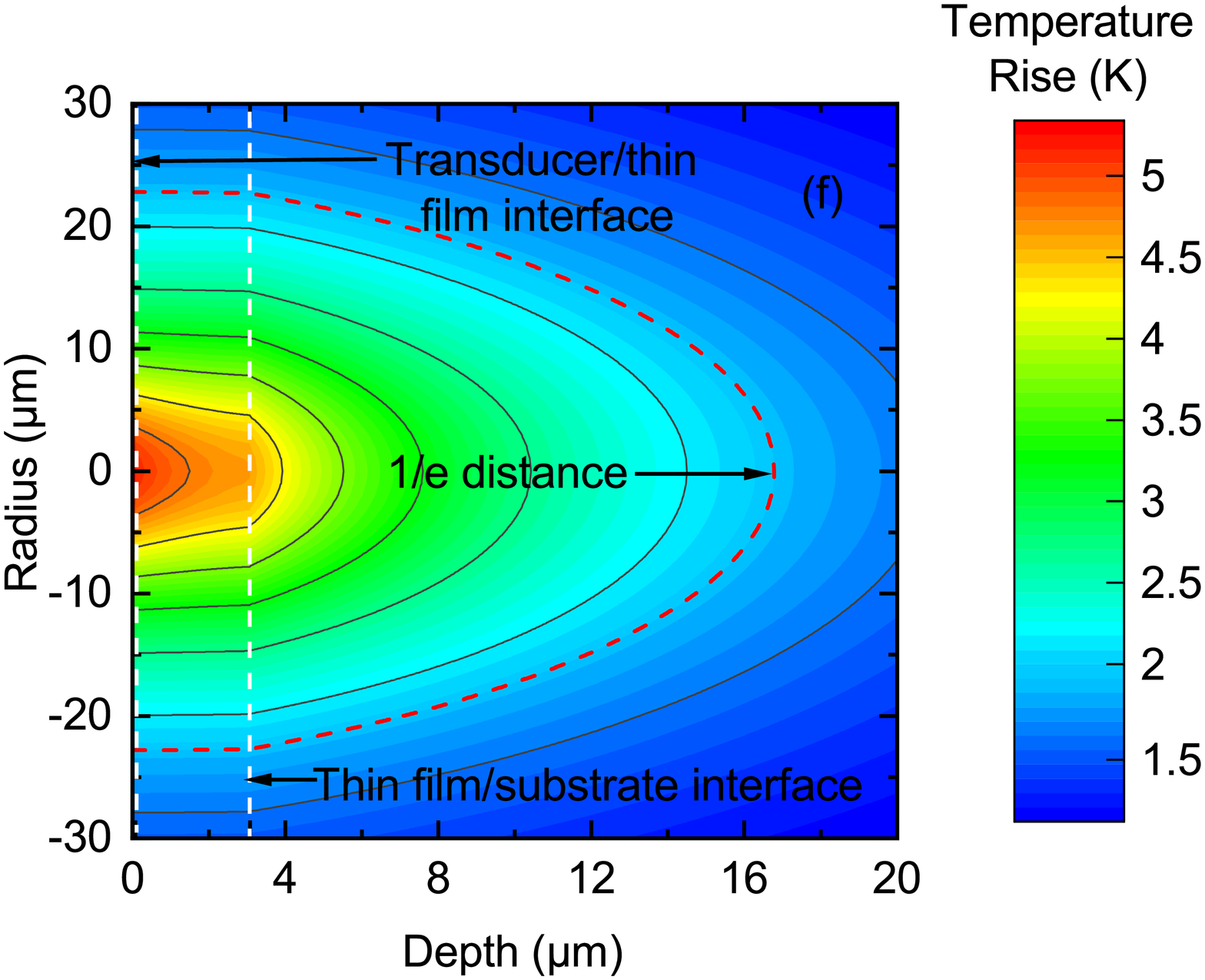}\hfill
\caption{ [(a) and (b)] Thermal penetration depth as a function of film thickness for a 3-layer system: metal transducer/thin film/substrate. [(c) and (d)] Normalized temperature drop as a function of depth for five different thin film thicknesses. [(e) and (f)] Temperature profiles of SSTR measurements for a 3 $\mu$m thin film on a substrate corresponding to an absorbed power of 5 mW. Figures (a), (c) and (e) represent the case of an insulating film on a conductive substrate (\textit{k}$_{2}$ = 10 W m$^{-1}$ K$^{-1}$ and \textit{k}$_{3}$ = 100 W m$^{-1}$ K$^{-1}$), whereas Figures (b), (d) and (f) represent the case of a conductive film on an insulating substrate (\textit{k}$_{2}$ = 100 W m$^{-1}$ K$^{-1}$ and \textit{k}$_{3}$ = 10 W m$^{-1}$ K$^{-1}$). The calculations correspond to \textit{f} = 100 Hz, \textit{d}$_1$ = 80 nm, \textit{r}$_0$ = \textit{r}$_1$ = 10 $\mu$m, \textit{k}$_{1}$ = 100 , \textit{C}$_{V, 1}$ = \textit{C}$_{V, 2}$ = \textit{C}$_{V, 3}$ = 2 MJ m$^{-3}$ K$^{-1}$, and \textit{G}$_{1}$ = \textit{G}$_{2}$ = 200 MW m$^{-2}$ K$^{-1}$.}
\label{fig:3}
\end{figure*}

We now extend the TPD discussion to a 3-layer system with the following geometry: metal transducer/thin film/substrate. When the thin film and substrate thermal conductivities are nearly equal, the TPD will closely follow those shown in Figure 2(a) with a minor influence from the thermal boundary conductance between the thin film and substrate. Thus, we consider two extreme cases of this hypothetical geometry: an insulating film on a conductive substrate (\textit{k}$_{2}$ = 10 W m$^{-1}$ K$^{-1}$, and \textit{k}$_{3}$ = 100 W m$^{-1}$ K$^{-1}$), and a conductive film on an insulating substrate (\textit{k}$_{2}$ = 100 W m$^{-1}$ K$^{-1}$, and \textit{k}$_{3}$ = 10 W m$^{-1}$ K$^{-1}$). 

In Figures 3(a) and 3(b), we present the TPD corresponding to the 1/e temperature drop distance as a function of thin film thickness for the first and second case, respectively. It is evident that the TPD with and without presence of a transducer are nearly identical. This is due to the fact that the thin film thermal conductivities are 10 and 100 W m$^{-1}$ K$^{-1}$, respectively. As shown in Figure 2(a), for this range of thermal conductivities, the transducer does not have a significant impact on the TPD. Similar to the 2-layer system, the 1/e$^{2}$ distance is much higher than the 1/e distance in the 3-layer system. From Figures 3(a) and 3(b), it is also clear that the TPD changes greatly with the film thickness when there is a significant difference between thin film and substrate thermal conductivities. Interestingly, the influence of the thin film on the TPD does not subside until the film thickness is approximately 4 times the heater radius. To understand the rationale behind this, it is necessary to review how the ratio of thin film to substrate thermal conductivity influences the heat flow direction. 

In Figures 3(c) and 3(d), we study the normalized temperature drop as a function of depth for the first and second case, respectively. When the thin film is insulating and the substrate is conductive, the bulk of the heat flows along the cross-plane direction of the thin film. Due to this, a large temperature gradient exists in the thin film along the cross-plane direction as shown in Figure 3(c). Therefore, in this case, the TPD is much lower than the heater radius unless the thin film thickness is too high or too low. On the other hand, when the thin film is conductive and the substrate is insulating, the majority of the heat flows along the in-plane direction of the thin film. Thus,  the temperature gradient along the cross-plane direction of the thin film is quite small. As a result, the TPD can be much higher than the heater radius as evident in Figure 3(d). 

To provide a more visual representation of this, the temperature profiles of SSTR measurements\cite{braun2017upper} are shown for a 3 $\mu$m thin film corresponding to the first and second case in Figures 3(e) and 3(f), respectively. For the insulating thin film case, temperature decreases greatly along the cross-plane direction of the film, whereas for the conductive film case, such temperature decrease is much smaller. However, for the conductive thin film case, temperature decrease is significant along the in-plane direction. This is in alignment with our previous discussion.

\subsection{Experimental verification of the thermal penetration depth definition}

Thus far, we have numerically predicted the TPD of 2-layer and 3-layer systems according to the conventional definition. We now conduct a series of experiments to check the validity of this conventional TPD definition for SSTR measurements. Specifically, we address the following questions: i) can SSTR probe up to the 1/e temperature drop distance defined by the traditional TPD description, and ii) whether this 1/e distance or the heater radius represents the absolute upper limit of how deep beneath the surface SSTR can probe.

\begin{figure*}[hbt!]
\centering
\includegraphics[scale=0.35]{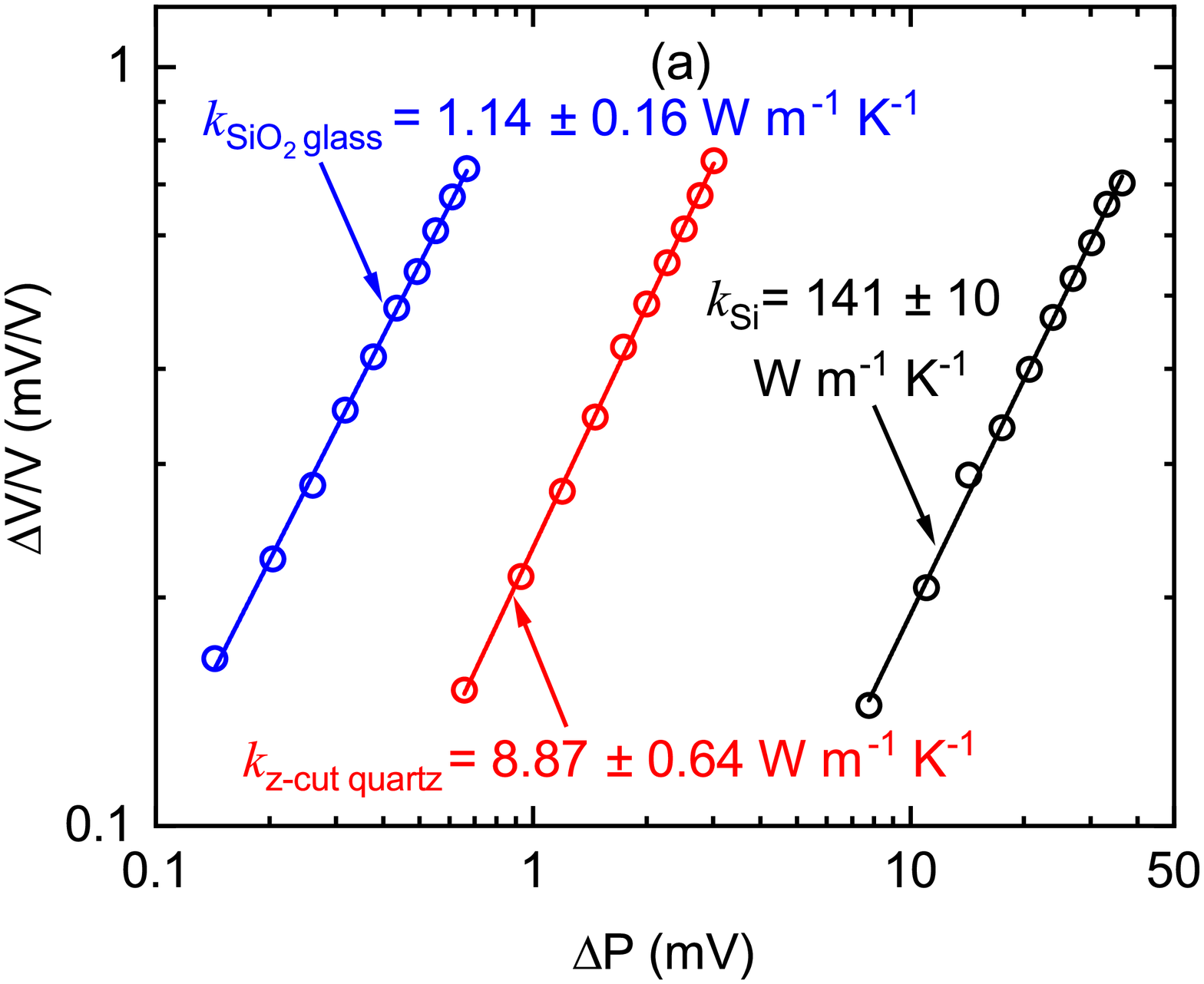} \hfill
\includegraphics[scale=0.35]{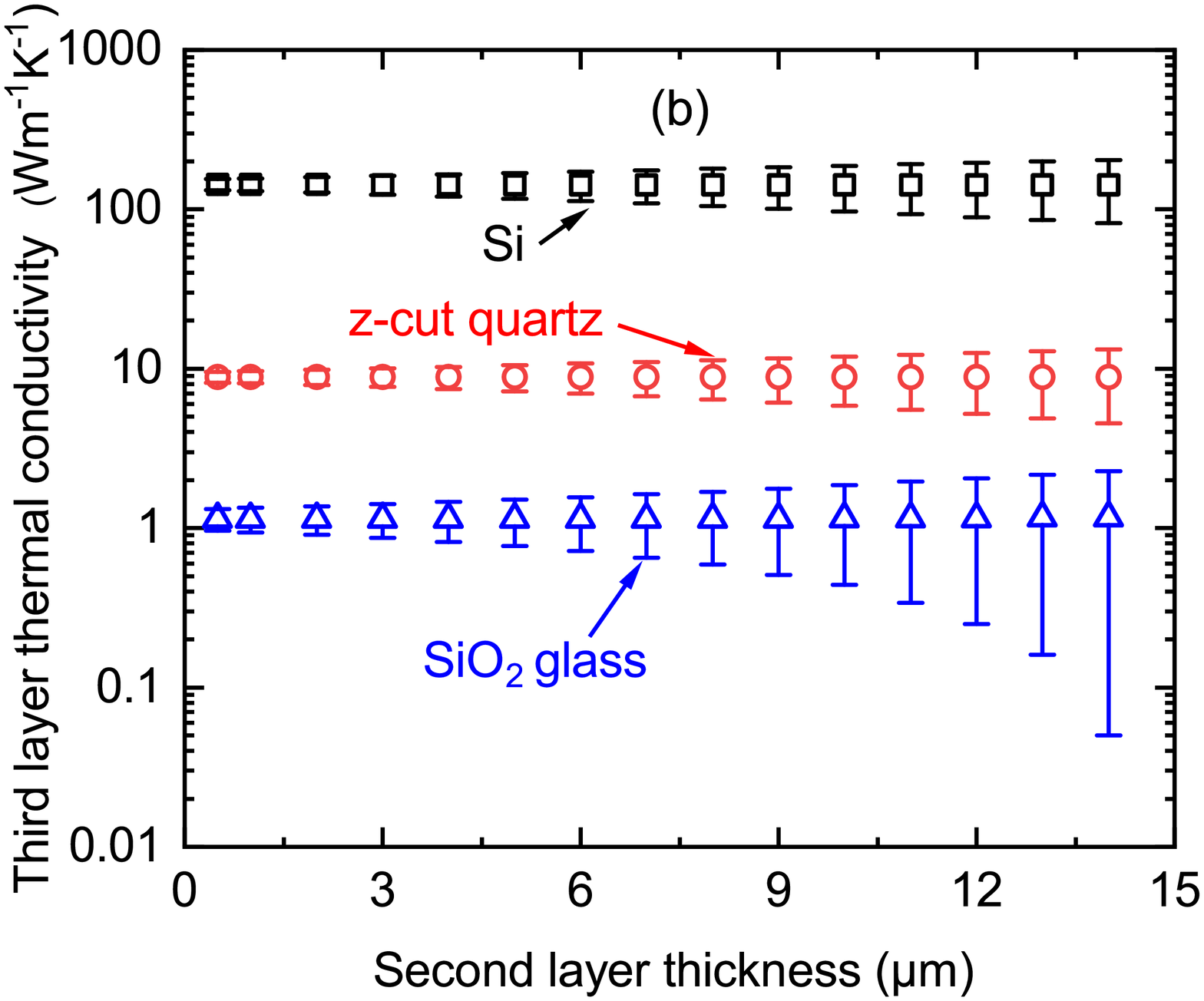} \\
\caption{(a) Probe photodetector response, $\triangle$\textit{V}/\textit{V} ($\propto$ temperature rise) as a function of pump photodetector response, $\triangle$\textit{P} ($\propto$ pump power) for SSTR fitting of Al coated bulk SiO$_2$ glass, z-cut quartz and Si. (b) Third layer thermal conductivity as a function of second layer thickness when the three samples are fitted as a 3-layer system: Al transducer/second layer/third layer. }
\label{fig:4}
\end{figure*}

For this purpose, the thermal conductivities ($\sqrt{k_rk_z}$) of three bulk samples are measured by SSTR: SiO$_{2}$ glass, z-cut quartz and Si. Prior to the measurements, the samples are coated with an $\sim$80 nm aluminum (Al) film to serve as an optical transducer. A pump radius of $\sim$10 $\mu$m is used for these SSTR measurements. The SSTR experimental proportionality constant, $\gamma$,\cite{braun2019steady} is determined from a reference sapphire sample (35 $\pm$ 2 W m$^{-1}$ K$^{-1}$).\cite{qin2020dual,jang2020thermoelectric} Details of our SSTR setup and measurement procedure have been thoroughly discussed in previous publications.\cite{braun2019steady} The 1/e temperature drop distance of the SiO$_{2}$ glass, z-cut quartz and Si samples are $\sim$10, 9 and 7 $\mu$m, respectively, according to the conventional TPD definition presented in Figure 2(a).

The SSTR best-fit curves for the thermal conductivities of the three samples are shown in Figure 4(a). The SSTR-measured thermal conductivities of the SiO$_2$ glass, z-cut quartz and Si are $\sim$1.14 $\pm$ 0.16, 8.87 $\pm$ 0.64, and 141 $\pm$ 10 W m$^{-1}$ K$^{-1}$, respectively. The uncertainty of the measured values stem from the uncertainty associated with the $\gamma$ value (sapphire reference), Al transducer thermal conductivity and the thermal boundary conductance. Details of these parameters are listed in Table 1. The measured thermal conductivities of the three specimen are in agreement with literature.\cite{fulkerson1968thermal,kremer2004thermal,wilson2014anisotropic,feser2014pump,braun2019steady}

\begin{table*}[b!]
\large
\centering
\begin{threeparttable}
\renewcommand{\tablename}{\large Table}
\renewcommand{\thetable}{\large 1}
\renewcommand{\arraystretch}{1.2}
\caption{\large Parameters used in the SSTR measurements and sensitivity calculations}
\begin{tabular}{ccccc}
\hline
\hline
\multirow{1}{*}{Samples} & \multirow{1}{*}{Layers\tnote{\color {blue} {a}}} & \multirow{1}{*}{Thermal conductivity\tnote{\color {blue} {b}}} & \multicolumn{2}{c}{Thermal boundary conductance\tnote{\color {blue} {c}}} \\
& & (W m$^{-1}$ K$^{-1}$) & \multicolumn{2}{c}{(MW m$^{-2}$ K$^{-1}$)}\\
\cline{4-5}
 &  &  & \textit{G}$_{1}$ & \textit{G}$_{2}$\\
\hline
Al/SiO$_{2}$ glass & Al & 126 $\pm$ 13 & 150 $\pm$ 20 & - \\
\hline
Al/Quartz & Al & 108 $\pm$ 12 & 230 $\pm$ 30 & - \\
\hline
Al/Si & Al & 117 $\pm$ 12 & 180 $\pm$ 30 & - \\
\hline
\multirow{2}{*}{Al/130 nm SiO$_{2}$/Si} & Al & 180 $\pm$ 18 & 230 $\pm$ 50 & -\\
\cline{2-5}
& SiO$_{2}$\tnote{\color {blue} {d}} & 1.35 $\pm$ 0.11 & - & 230 $\pm$ 50 \\
\hline
\multirow{2}{*}{Al/2.05 $\mu$m GaN/GaN} & Al & 130 $\pm$ 13 & 240 $\pm$ 40 & -\\
\cline{2-5}
& GaN & 184 $\pm$ 15 & - & 150 $\pm$ 50\\
\hline
\multirow{2}{*}{Al/2 $\mu$m AlN/sapphire} & Al & 190 $\pm$ 19 & 380 $\pm$ 80 & -\\
\cline{2-5}
& AlN & 281 $\pm$ 26 & - &  150 $\pm$ 50\\
\hline
\multirow{2}{*}{Al/2.5 $\mu$m Si/1 $\mu$m SiO$_{2}$/Si} & Al & 180 $\pm$ 18 & 100 $\pm$ 10 & -\\
\cline{2-5}
& Si & 127 $\pm$ 11 & - & 230 $\pm$ 50\tnote{\color {blue} {e}}\\
\hline
\hline
\end{tabular}
\begin{tablenotes}
    \item[\color {blue} {a}] The thicknesses of the layers are measured by picosecond acoustics and transmission elctron microscopy (TEM). The uncertainty associated with layer thicknesses are about $\sim$2-3$\%$.
    \item[\color {blue} {b}] The thermal conductivities of the Al transducers are measured by 4-point probe. The SiO$_{2}$, GaN, AlN and Si thin film thermal conductivities are measured by TDTR.
    \item[\color {blue} {c}] \textit{G}$_{1}$ is measured by TDTR. \textit{G}$_{2}$ is estimated from related literature.\cite{ziade2015thermal,braun2016size,tareq2019investigation,cheng2020interfacial,giri2020review} As SSTR has negligible sensitivity to \textit{G}$_{2}$, the estimated values do not have an appreciable influence on SSTR measurements.
    \item[\color {blue} {d}] In the Al/SiO$_{2}$ glass sample, the SiO$_{2}$ substrate is a commercial glass slide, whereas in Al/130 nm SiO$_{2}$/Si sample, the SiO$_{2}$ thin film is laboratory grade SiO$_{2}$ grown via dry oxidation.\cite{braun2016size} As a result, the thermal conductivity of SiO$_{2}$ is different between the two samples.
    \item[\color {blue} {e}] For the Al/2.5 $\mu$m Si/1 $\mu$m SiO$_{2}$/Si sample, the thermal boundary conductances of the Si/SiO$_{2}$ (\textit{G}$_{2}$) and SiO$_{2}$/Si (\textit{G}$_{3}$) interfaces are considered to be the same.
  \end{tablenotes}
\end{threeparttable}
\end{table*}

In Figure 4(b), the SiO$_2$ glass, z-cut quartz and Si samples are approximated as a 3-layer material system: Al transducer/second layer/third layer. Here, the second and third layer represent thin films and buried substrates, respectively. We fit for the thermal conductivity of the third layer assuming that the second layer possesses the value presented in Figure 4(a). The thermal boundary conductance between the second and third layer is kept fixed at 1000 MW m$^{-2}$ K$^{-1}$. Figure 4(b) shows that with the increase of second layer thickness, the uncertainty of third layer thermal conductivity increases. When the second layer thickness is equal to the 1/e distance ($\sim$10, 9 and 7 $\mu$m for SiO$_{2}$ glass, z-cut quartz and Si, respectively), the third layer thermal conductivities are $\sim$1.15 $\pm$ 0.71, 8.88 $\pm$ 2.74, and 142.5 $\pm$ 33.7 W m$^{-1}$ K$^{-1}$, respectively. Furthermore, when the second layer thickness is 14 $\mu$m, the third layer thermal conductivities are $\sim$1.16 $\pm$ 1.11, 8.89 $\pm$ 4.35, and 142.7 $\pm$ 61 W m$^{-1}$ K$^{-1}$, respectively.

It is possible to answer the previously posed questions from Figure 4(b). As shown here, SSTR can measure the thermal conductivities of layers located at 1/e temperature drop distance although such measurements have relatively high uncertainty. However, it is also evident that SSTR can probe beyond this conventional 1/e distance and the heater radius. This indicates that immediately beyond the 1/e distance or the heater radius, SSTR measurement sensitivity does not drop to zero. This phenomenon can be explained by reviewing Figure 2(a) which shows that the temperature does not drop to 1/e$^{2}$ value of maximum surface temperature until the distance is much higher than  the 1/e distance or the heater radius. Thus, even though the traditional TPD definition can be used as a convenient estimate of SSTR probing depth, neither the 1/e distance nor the heater radius should be taken as an absolute upper limit of how far beneath the surface SSTR can probe. 

\subsection{Influence of multilayer material systems on the thermal conductivity measurements of buried substrates}

To empirically study how different parameters of multilayer material systems impact the thermal conductivity measurements of buried layers or substrates, we use the same example used in section 2.3. Figure 5(a) shows the $\%$ uncertainty of the third layer thermal conductivity as a function of second layer thickness corresponding to Figure 4(b). The uncertainty of the third layer thermal conductivity is highest for SiO$_{2}$ glass, followed by z-cut quartz and Si. This may seem counter intuitive as SiO$_{2}$ glass has the highest 1/e temperature drop distance among the three materials. Thus, one might expect the third layer of the SiO$_{2}$ glass to have the lowest uncertainty among the samples. To understand this apparent anomaly, it is necessary to review how sensitivity to different parameters influence the SSTR measurements of third layer thermal conductivity.   

\begin{figure*}[hbt!]
\centering
\includegraphics[scale=0.35]{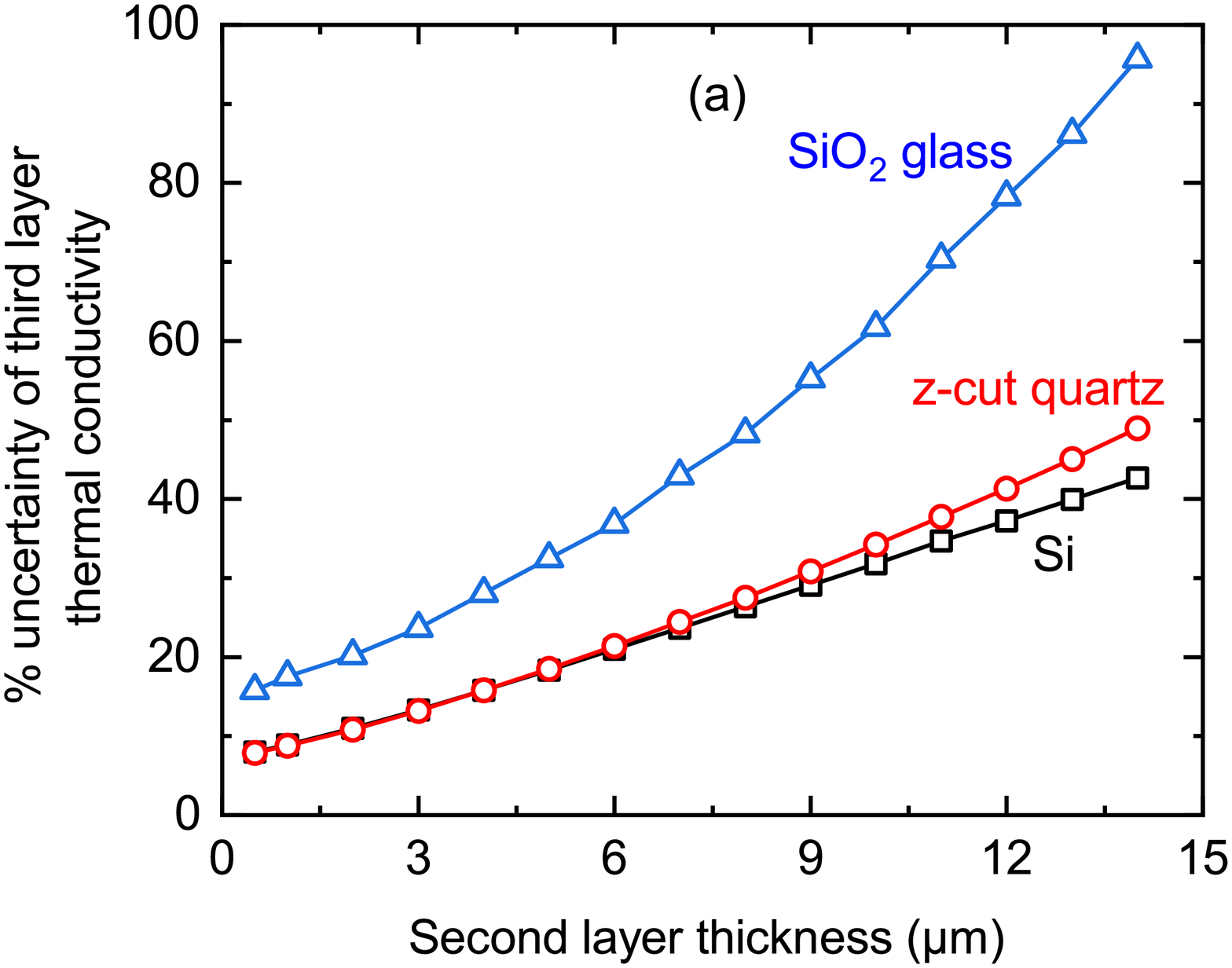} \\
\includegraphics[scale=0.21]{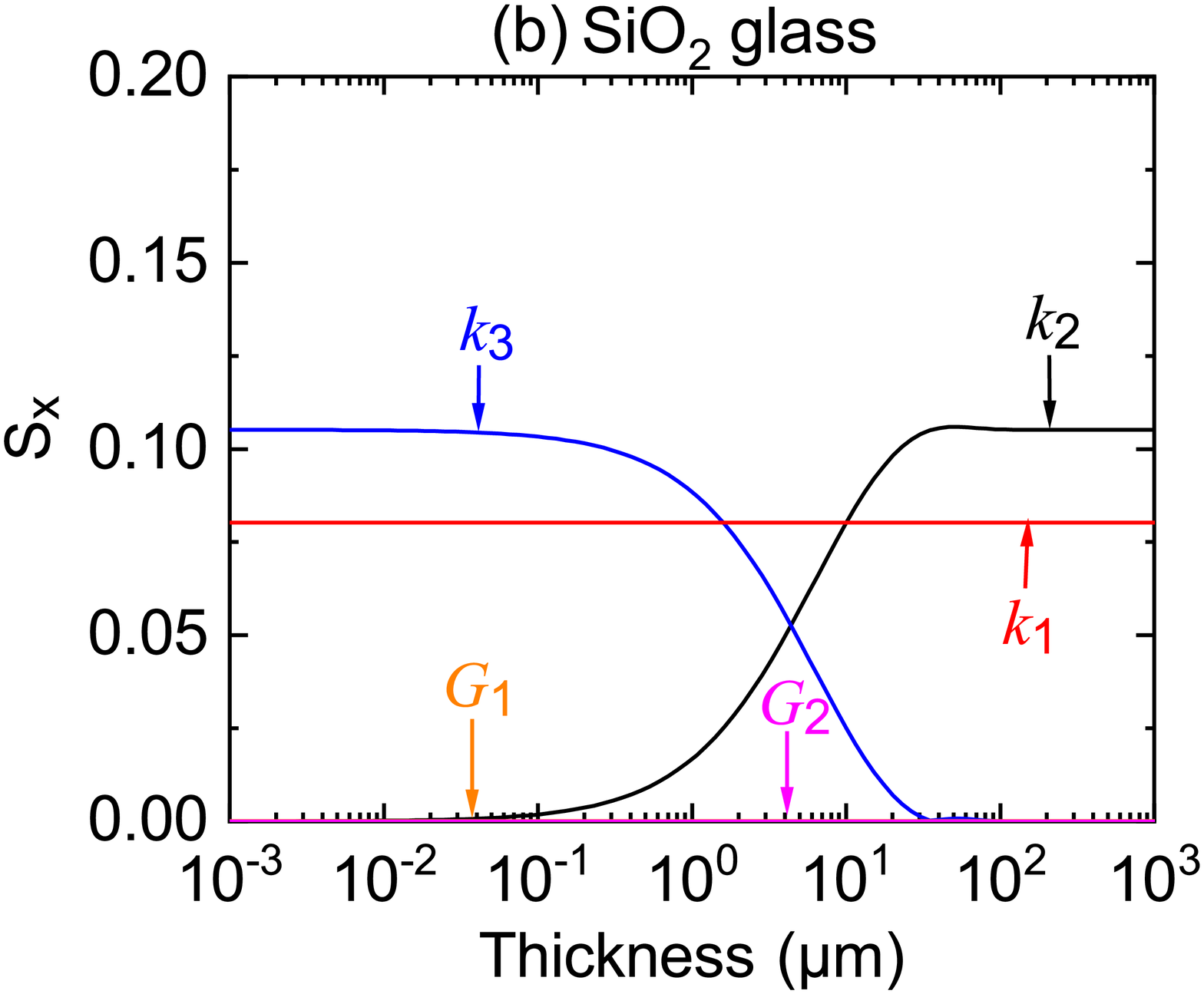} \hfill
\includegraphics[scale=0.21]{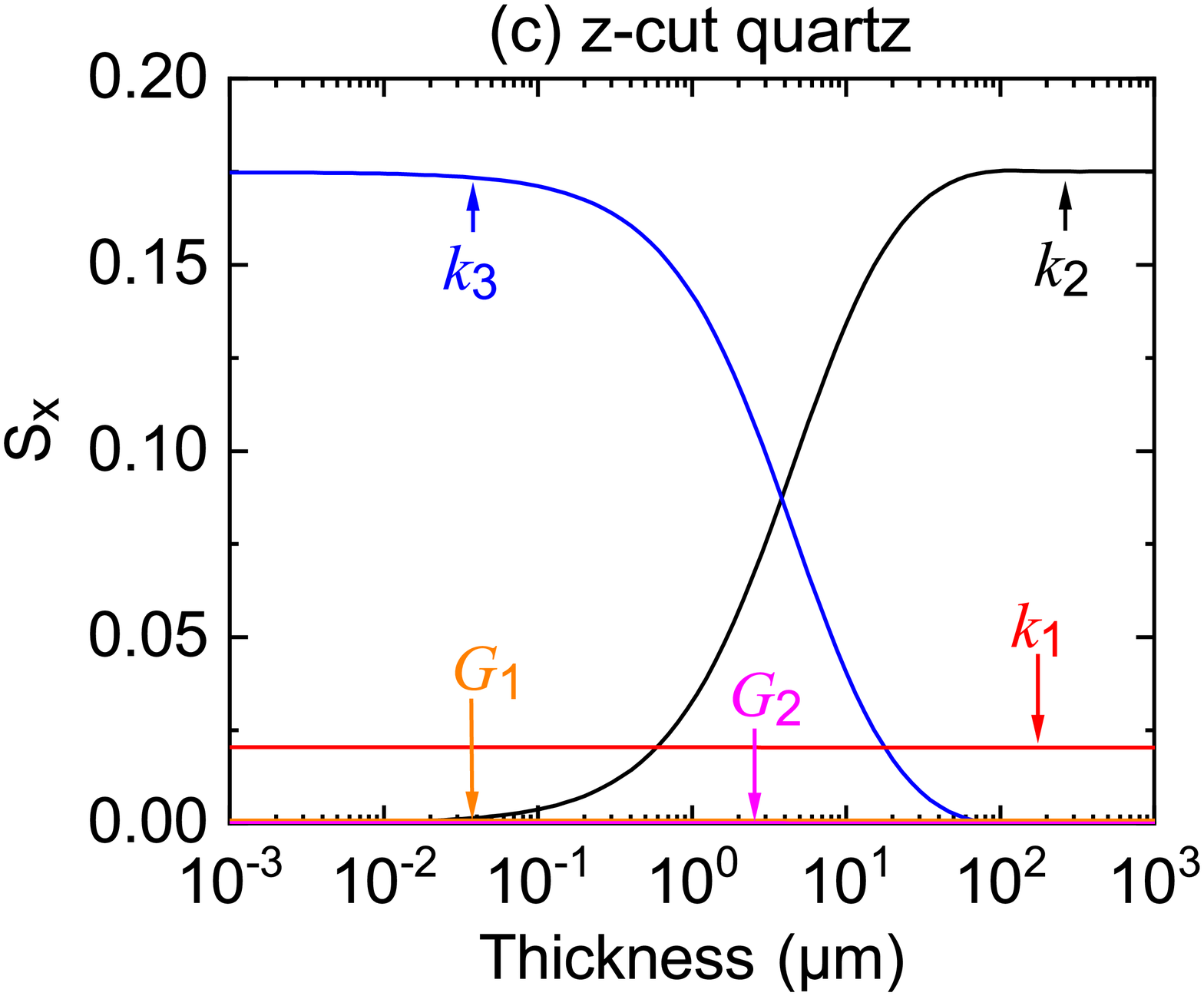} \hfill
\includegraphics[scale=0.21]{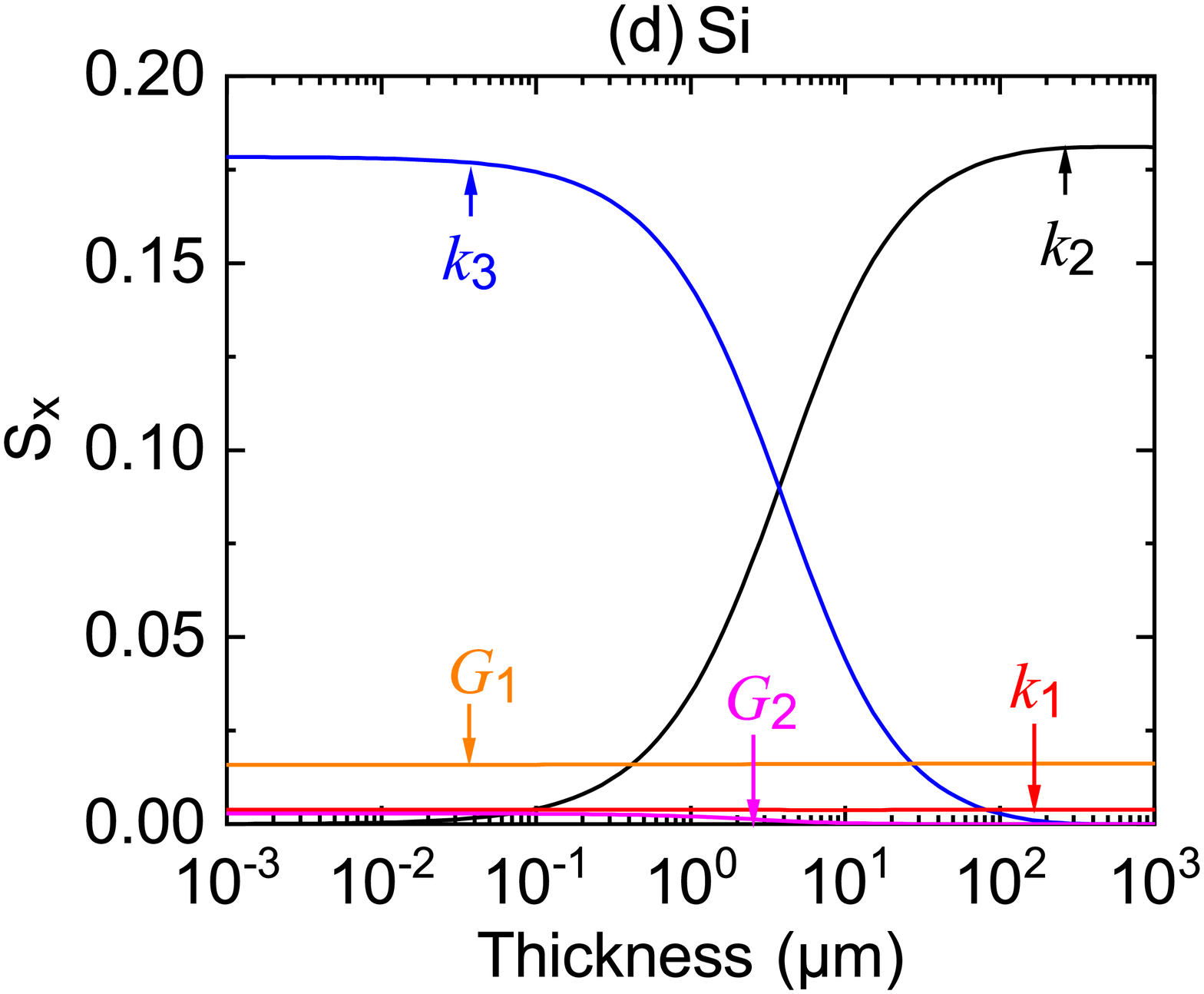} \hfill
\caption{(a) $\%$ uncertainty of the third layer thermal conductivity as a function of second layer thickness correspoding to Figure 4(b). The sensitivity, S$_{x}$, as a function of second layer thickness for (d) SiO$_{2}$ glass, (e) z-cut quartz, and (f) Si. For all specimen, \textit{k} represents $\sqrt{k_rk_z}$.}
\label{fig:4}
\end{figure*}

The sensitivity of SSTR measurements to different parameters for SiO$_{2}$ glass, z-cut quartz and Si are presented in Figures 5(b), 5(c) and 5(d), respectively. It is evident from these sensitivity calculations that when the second layer thickness is high, measurements of third layer are greatly impacted by the second layer thermal conductivity. However, for SiO$_2$ glass, there is also significant sensitivity to the transducer thermal conductivity, whereas for z-cut quartz and Si, sensitivity to nearly all other parameters are very small. Due to the influence of second layer and transducer thermal conductivity, the uncertainty of SiO$_2$ glass is highest. For similar reasons, the uncertainty of the third layer thermal conductivity is higher for z-cut quartz compared to Si when the second layer thickness is high. At such thicknesses, SSTR measurements of z-cut quartz become sensitive to the transducer thermal conductivity. Although Si measurements also become sensitive to the Al/Si interface conductance, the sensitivity of Si to this thermal boundary conductance is lower compared to the sensitivity of z-cut quartz to the transducer thermal conductivity. As a result, the uncertainty of Si measurements are relatively lower than z-cut quartz at high second layer thicknesses.   

From the above discussion, it can be concluded that the TPD cannot provide an estimation of the uncertainty associated with SSTR measurements of a buried substrate. Instead, such uncertainty depends on how sensitive SSTR measurements are to different parameters such as the transducer thermal conductivity, the thermal boundary conductances, and the thermal resistances of different layers of the multilayer material system. Therefore, sensitivity calculations can provide the best means for estimating the uncertainty of a buried layer or substrate thermal conductivity.

\subsection{Sensitivity calculations for SSTR measurements of buried substrates using different heater radii}

Figure 6 shows the sensitivity calculations for a 3-layer system: metal transducer/thin film/substrate for three different heater radii: 2, 20 and 50 $\mu$m. SSTR measurements are most sensitive to the substrate thermal conductivity when the heater radius is much larger than the thin film thickness, irregardless of what the thin film to substrate thermal conductivity ratio is. When the heater radius is small (i.e., 2 $\mu$m), the sensitivity to the in-plane and cross-plane thermal conductivity of the substrate are nearly the same. However, as the heater radius increases, sensitivity to the in-plane thermal conductivity of the substrate keeps decreasing. This occurs because larger heater radius requires longer time to reach steady-state. Therefore, to increase the sensitivity to the in-plane thermal conductivity of the substrate, the modulation frequency needs to be lowered. From Figure 6, it is evident that by changing the heater radius, it is possible to measure the thermal conductivity of buried substrates for different thin film thicknesses.  

\begin{figure*}[hbt!]
\includegraphics[scale=0.216]{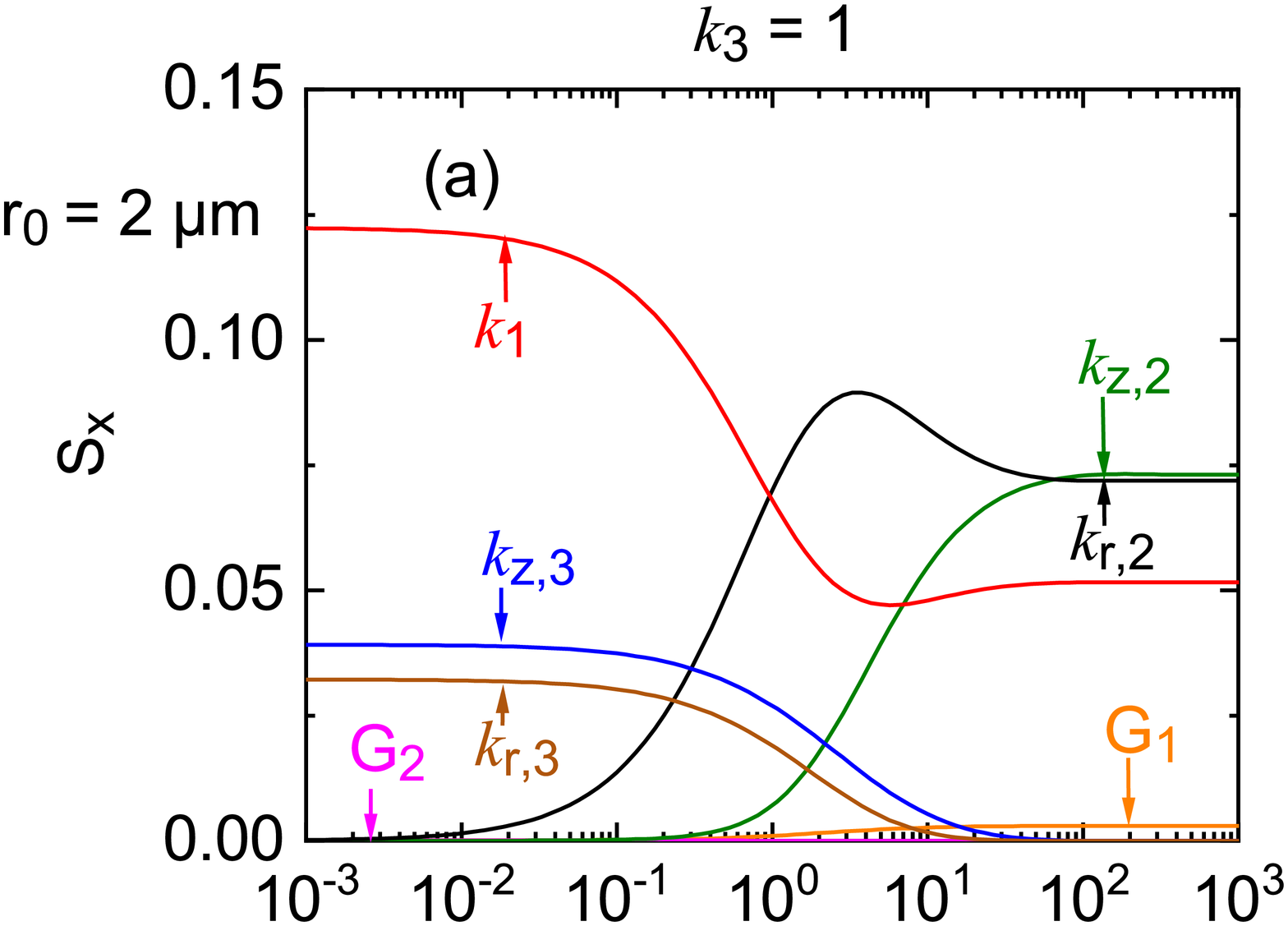} \hfill
\includegraphics[scale=0.216]{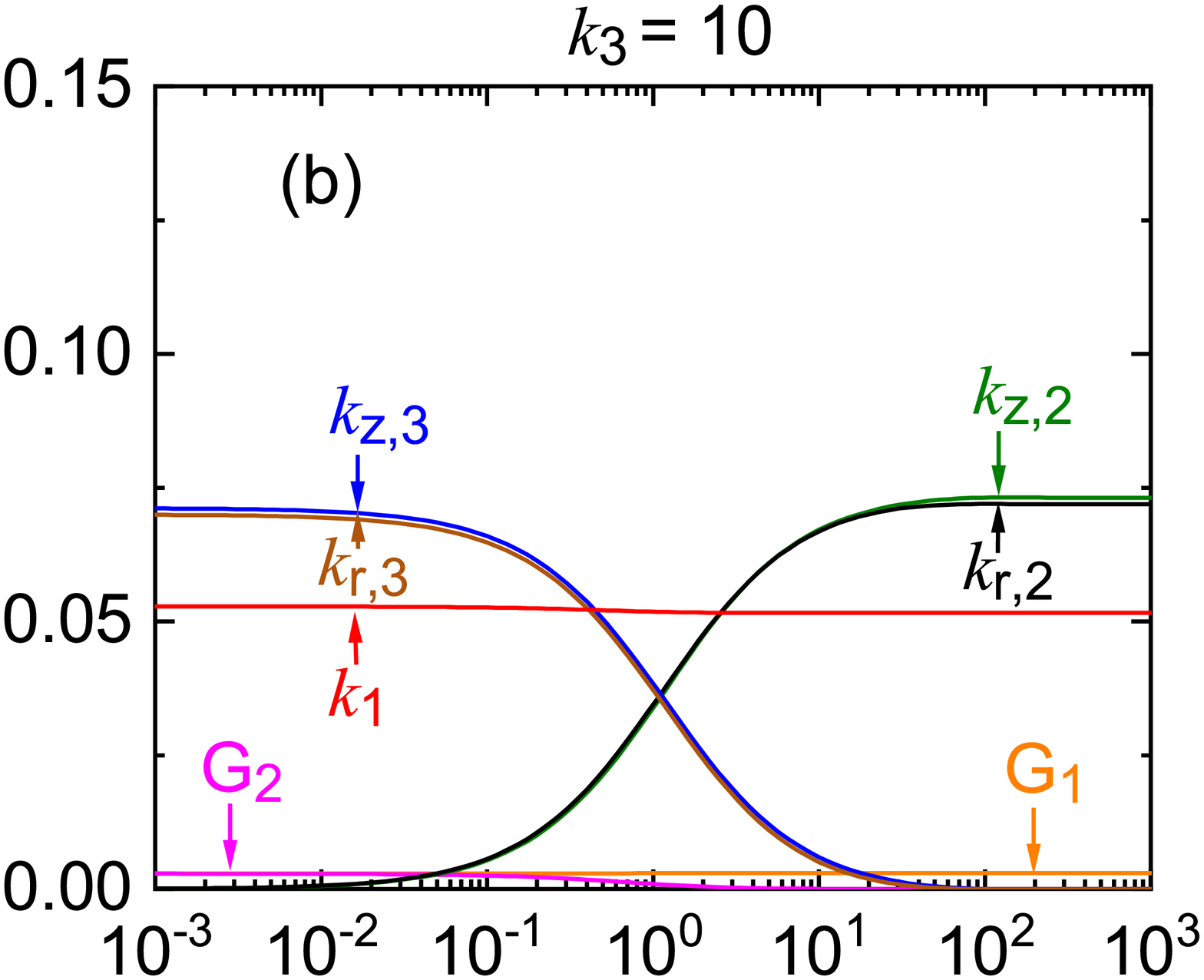} \hfill
\includegraphics[scale=0.216]{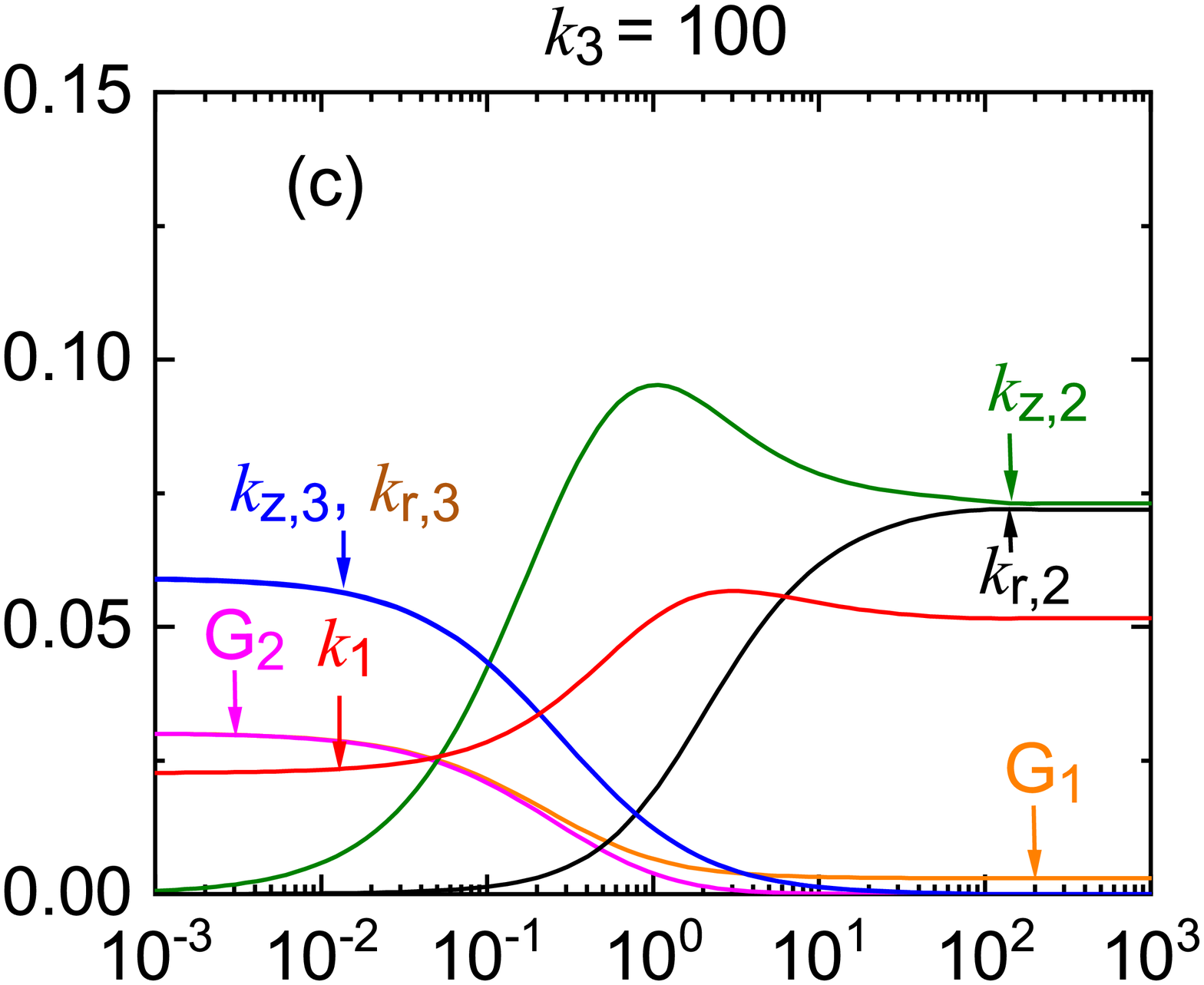} \\
\includegraphics[scale=0.216]{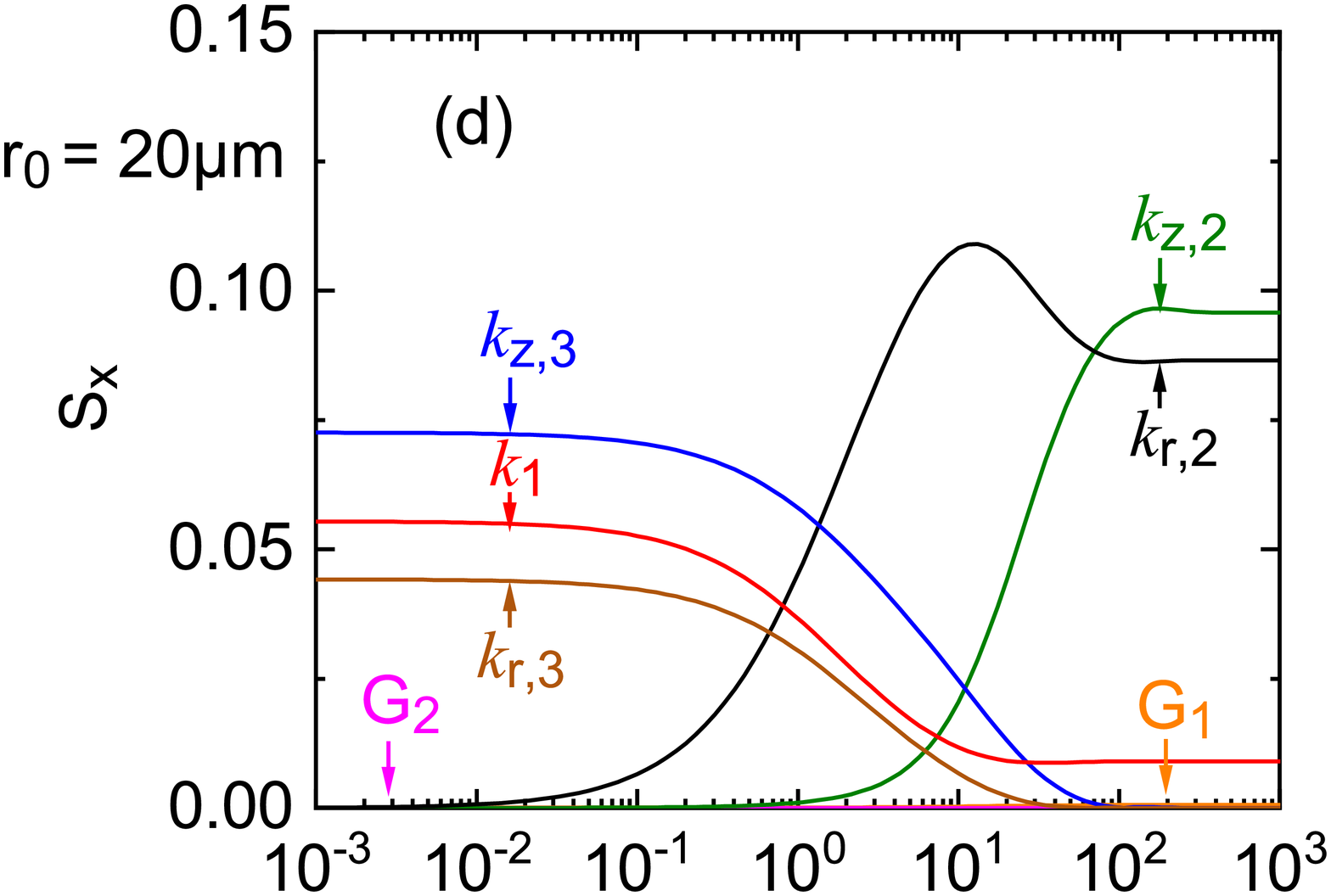} \hfill
\includegraphics[scale=0.216]{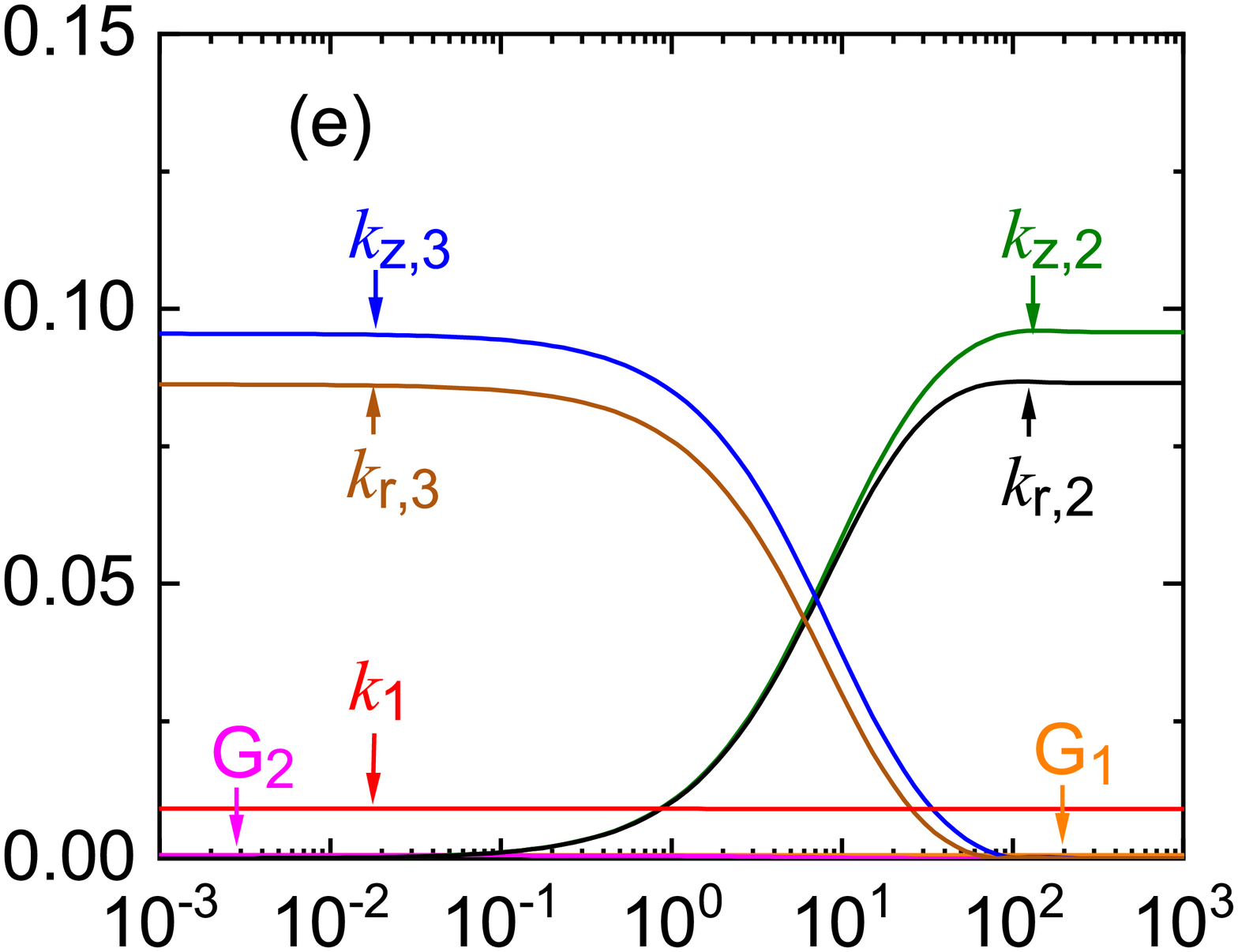} \hfill
\includegraphics[scale=0.216]{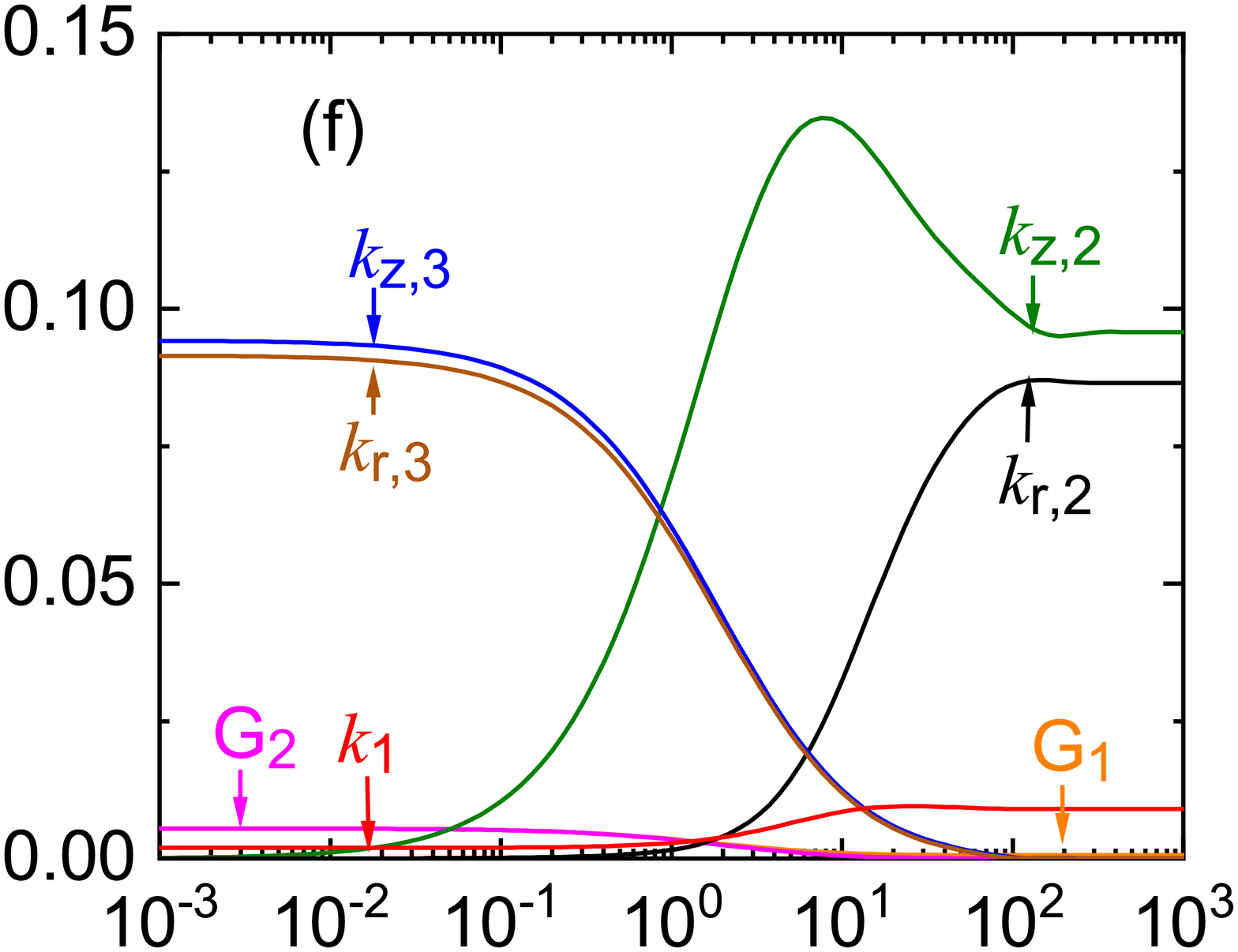} \\
\includegraphics[scale=0.216]{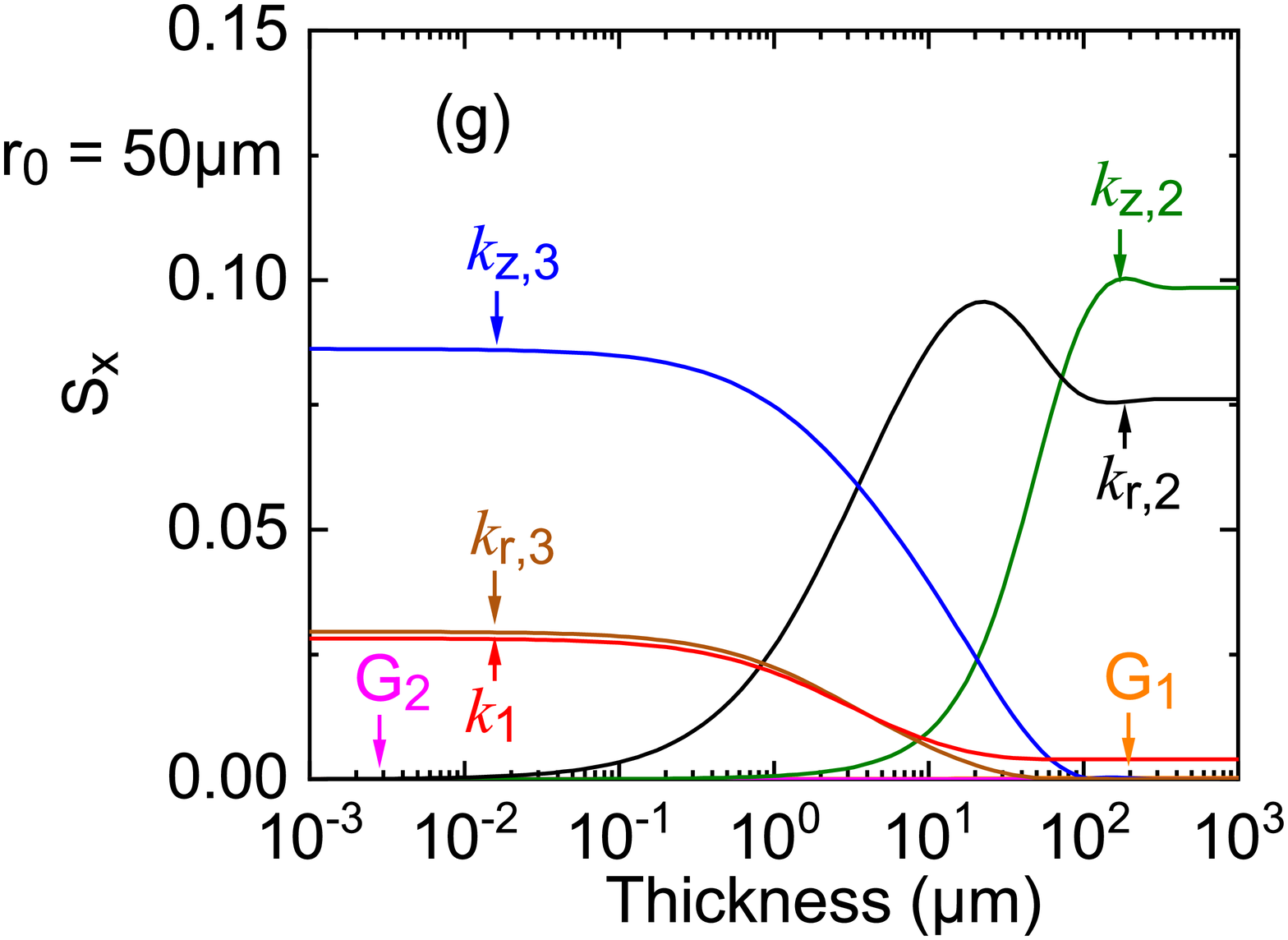} \hfill
\includegraphics[scale=0.216]{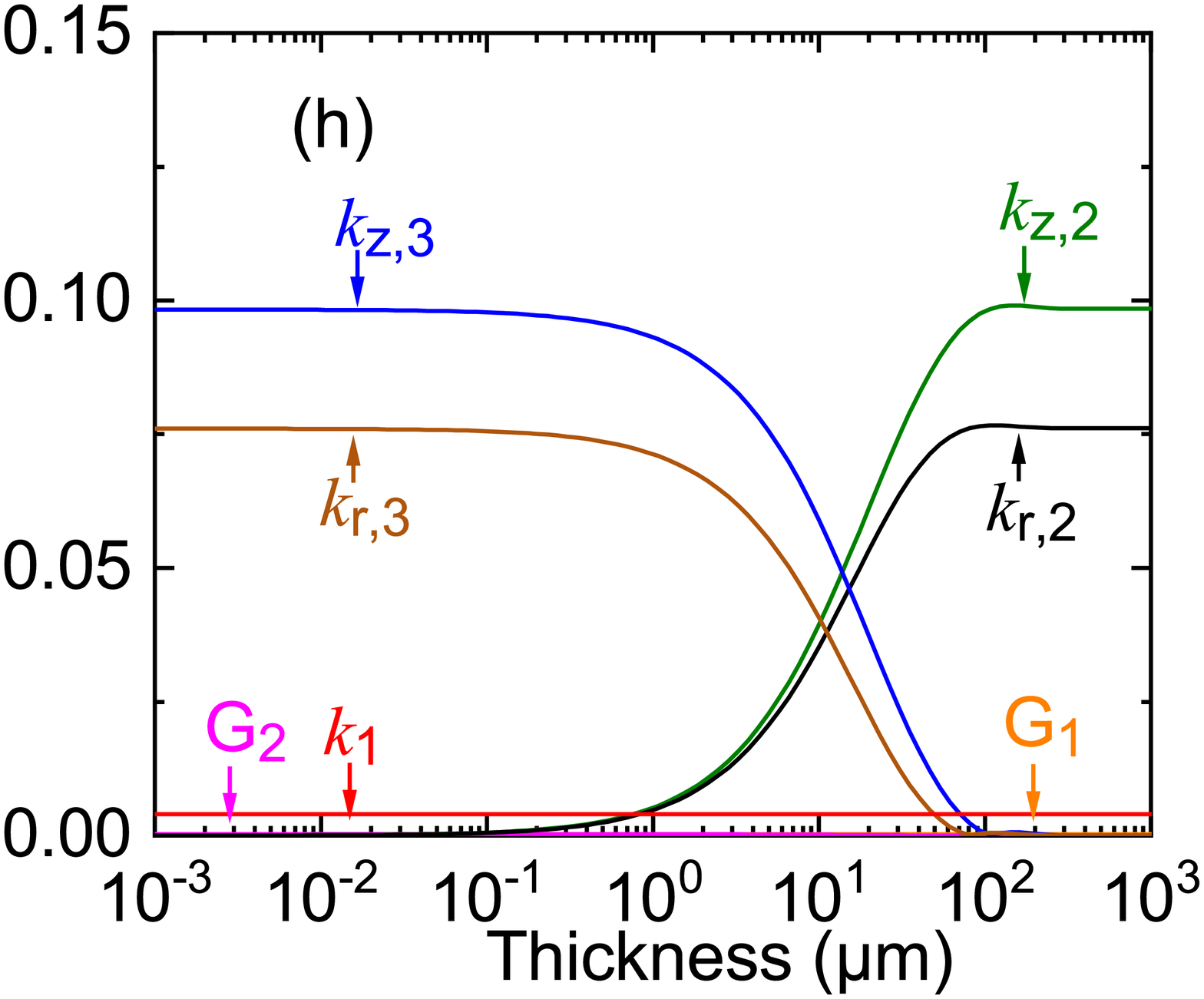} \hfill
\includegraphics[scale=0.216]{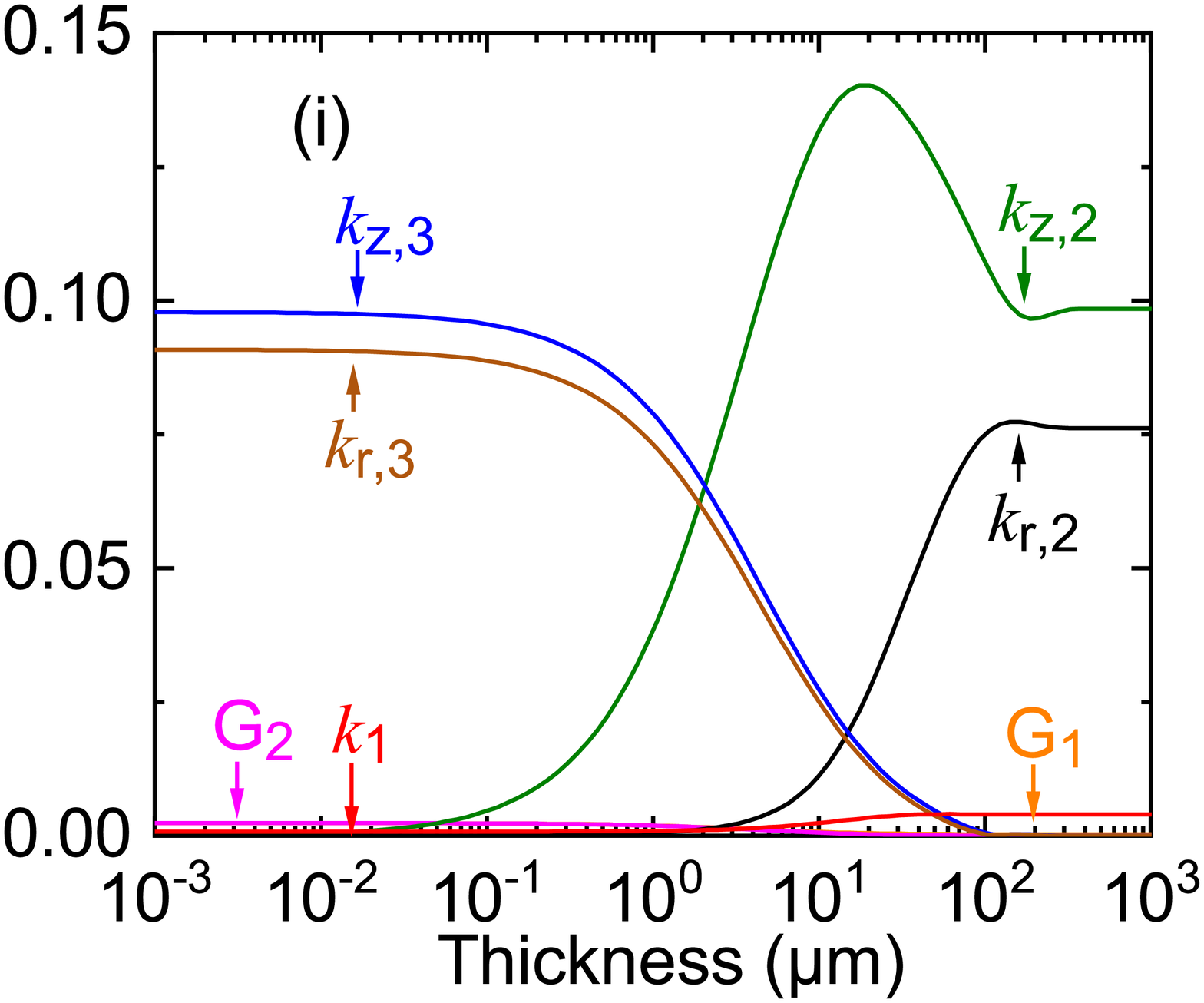} \hfill

\caption{(a) Sensitivity, S$_{x}$ as a function of thin film thickness for a 3-layer system: metal transducer/thin film/substrate. Three different heater radii (\textit{r}$_0$) are considered here: 2, 20 and 50 $\mu$m. The sensitivity calculations correspond to \textit{f} = 100 Hz, \textit{d}$_1$ = 80 nm, \textit{r}$_1$ = \textit{r}$_0$, \textit{k}$_{1}$ = 100 W m$^{-1}$ K$^{-1}$, \textit{C}$_{V, 1}$ = \textit{C}$_{V, 2}$ = \textit{C}$_{V, 3}$ = 2 MJ m$^{-3}$ K$^{-1}$, \textit{G}$_{1}$ = \textit{G}$_{2} = $200 MW m$^{-2}$ K$^{-1}$, and \textit{k}$_{2}$ = 10 W m$^{-1}$ K$^{-1}$.}
\label{fig:5}
\end{figure*}

\subsection{Experimental measurements of the thermal conductivity of buried substrates by SSTR}

To further demonstrate the ability of SSTR to measure the thermal conductivity ($\sqrt{k_rk_z}$) of buried substrates, we choose three samples with the following 3-layer geometry: Al transducer/thin film/substrate. The schematics of the three samples are shown in Figures 7(a)-(c). The first sample is a $\sim$130 nm a-SiO$_{2}$ thin film on Si substrate. This sample represents an insulating film on a conductive substrate. The second sample is a $\sim$2.05 $\mu$m unintentional doped (UID) GaN thin film on hydride vapor phase epitaxy (HVPE) n-GaN substrate. This sample represents the case where the thin film and substrate thermal conductivities are nearly equal. The third sample is a $\sim$2 $\mu$m molecular beam epitaxy (MBE) grown AlN thin film on sapphire substrate. This sample represents a conductive film on an insulating substrate. Traditional pump-probe techniques such as TDTR and FDTR often can not measure the thermal conductivity of buried substrates in such samples due to their limited thermal penetration depths under standard operating conditions. Moreover, for the Si substrate measurements, TDTR and FDTR can also suffer from non-equilibrium processes, interfacial phonon-scattering, and ballistic phonon transport.\cite{minnich2011thermal,regner2013broadband,wilson2014anisotropic,wilson2015limits}

\begin{figure*}[hbt!]
\centering
\includegraphics[scale=0.35]{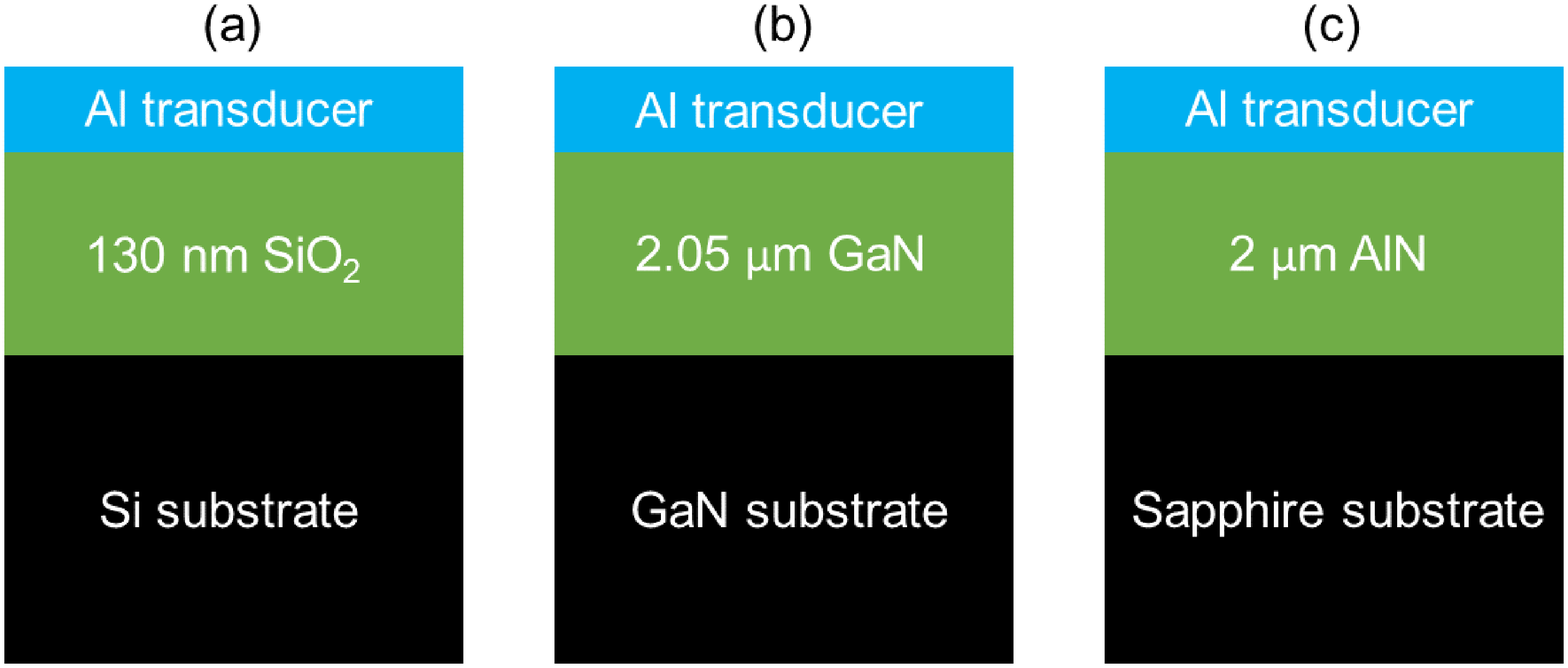} \\
\includegraphics[scale=0.21]{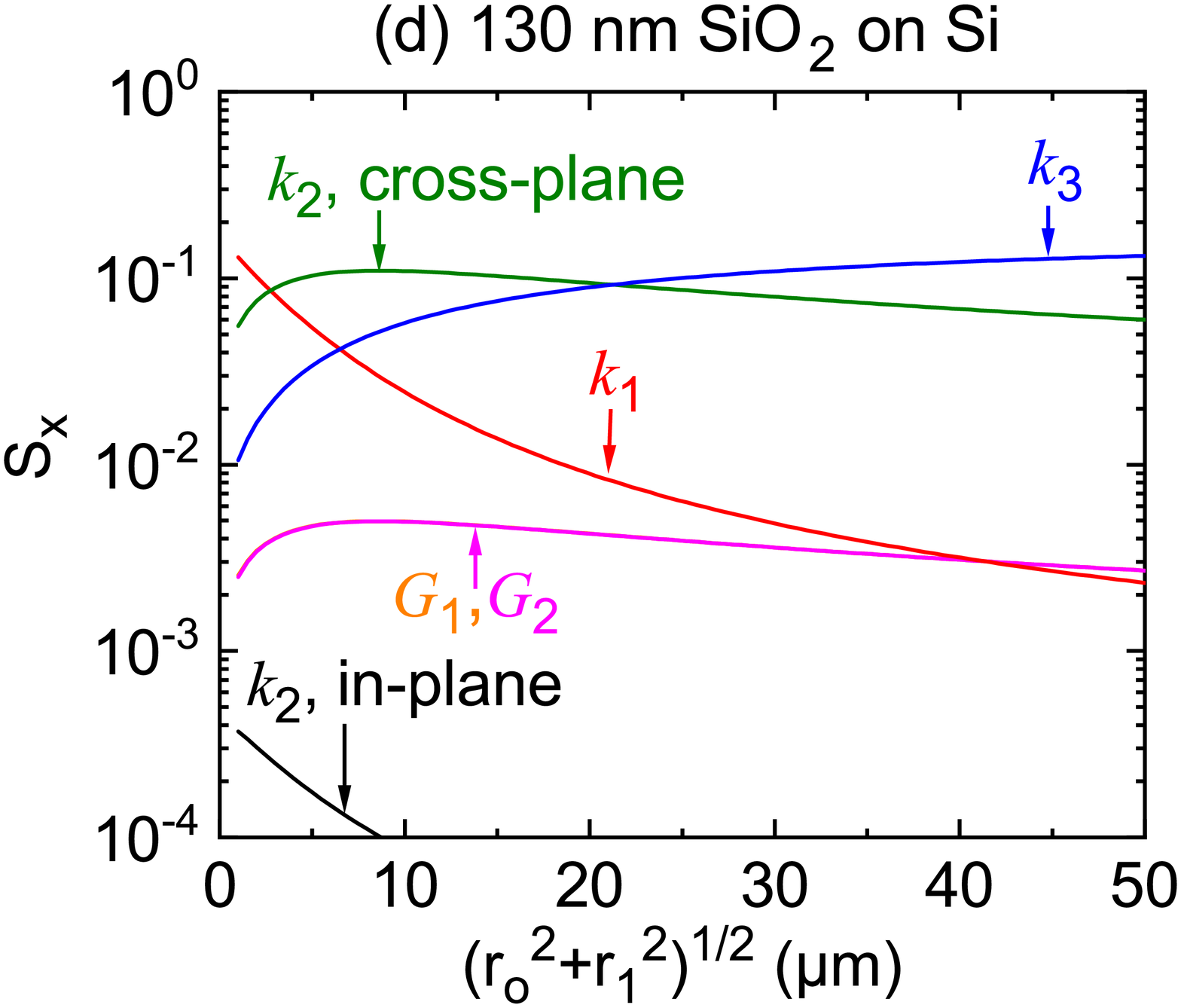} \hfill
\includegraphics[scale=0.21]{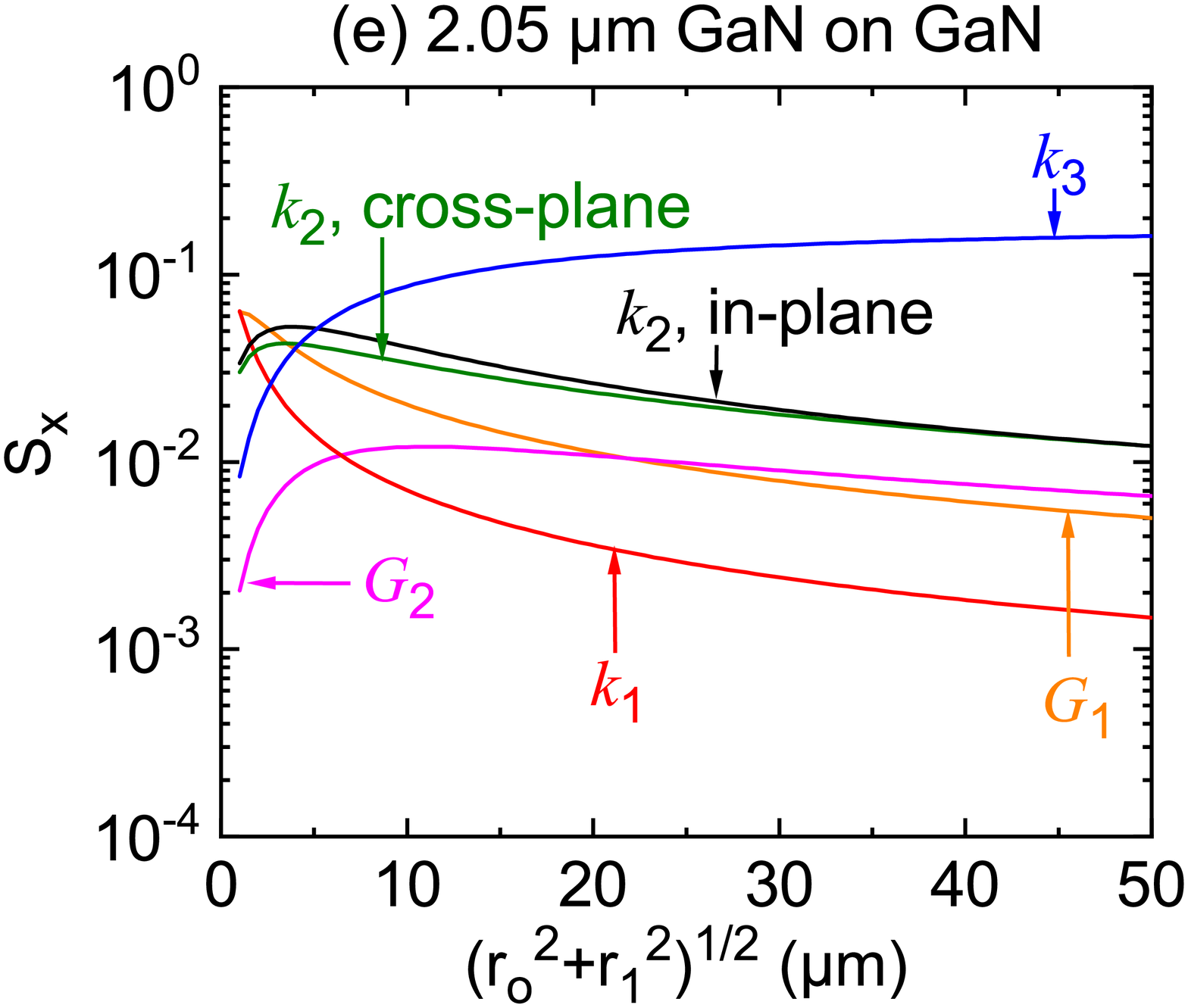} \hfill
\includegraphics[scale=0.21]{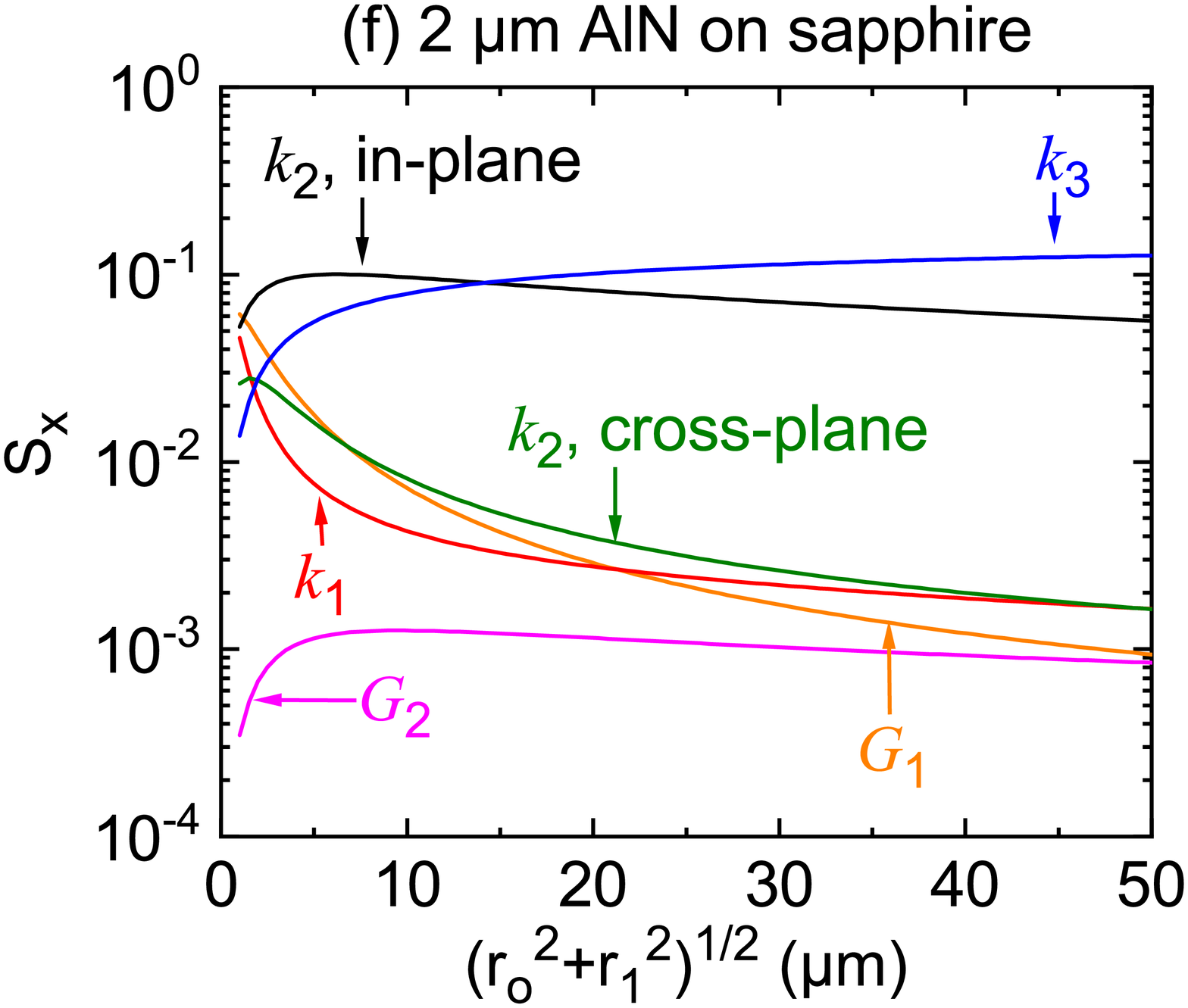} \hfill
\caption{Schematics of the 3-layer samples measured by SSTR: (a) $\sim$130 nm SiO$_{2}$ thin film on Si substrate, (b) $\sim$2.05 $\mu$m GaN thin film on n-GaN substrate, and (c) $\sim$2 $\mu$m AlN thin film on sapphire substrate. Figures (d), (e), and (f) represent the sensitivity calculations as a function of effective radius, $\sqrt{r{_o^2}+r{_1^2}}$ for the three samples shown in Figures (a), (b), and (c) respectively.}
\label{fig:6}
\end{figure*}

\begin{table*}
\large
\centering
\renewcommand{\tablename}{\large Table}
\renewcommand{\thetable}{\large 2}
\renewcommand{\arraystretch}{1.2}
\caption{\large SSTR-measured substrate thermal conductivity ($\sqrt{k_rk_z}$) of the samples shown in Figure 7}
\begin{tabular}{cccc}
\hline
\hline
\multirow{2}{*}{Substrates} & \multicolumn{3}{c}{Thermal conductivity (W m$^{-1}$ K$^{-1}$)} \\
\cline{2-4}
 & spot size 10 $\mu$m & spot size 20 $\mu$m & literature \\
\hline
Si & 141 $\pm$ 27 & 140 $\pm$ 18 & 140\cite{fulkerson1968thermal}\\
GaN & 194 $\pm$ 27 & 185 $\pm$ 16 & 195\cite{florescu2000high}\\
Sapphire & 35.1 $\pm$ 5.9 & 34.5 $\pm$ 4.2 & 35\cite{cahill1998thermal}\\
\hline
\end{tabular}
\end{table*}

In the $\sim$130 nm a-SiO$_{2}$ thin film on Si sample, we use co-axially focused 1/e$^{2}$ pump and probe radii of $\sim$10 $\mu$m to measure the thermal conductivity of buried Si substrate. The sensitivity calculation for this sample is shown in Figure 7(d). As shown here, SSTR measurements of the Si substrate are sensitive to the cross-plane thermal conductivity of the SiO$_{2}$ thin film. TDTR is used to measure the cross-plane thermal conductivity of the SiO$_{2}$ thin film. Using this SiO$_{2}$ value as an input, the SSTR-measured thermal conductivity of the Si substrate is 141 $\pm$ 27 W m$^{-1}$ K$^{-1}$. Figure 7(d) indicates that the sensitivity to SiO$_{2}$ cross-plane thermal conductivity is very high when the pump and probe radii are 10 $\mu$m. As a result, the uncertainty associated with the Si thermal conductivity is also high, $\sim$19$\%$. The sensitivity calculation also shows that using larger spot sizes the sensitivity to SiO$_{2}$ and corresponding uncertainty of Si measurement can be reduced. To demonstrate this, we repeat the measurement with 1/e$^{2}$ pump and probe radii of $\sim$20 $\mu$m. The resultant Si thermal conductivity is 140 $\pm$ 18 W m$^{-1}$ K$^{-1}$. As predicted, the measurement with the 20 $\mu$m spot sizes has a reduced uncertainty of $\sim$13$\%$.

In the $\sim$2.05 $\mu$m GaN thin film on n-GaN substrate sample, we measure the thermal conductivity of the n-GaN substrate by SSTR using $\sim$10 and 20 $\mu$m spot sizes. The sensitivity calculation for this sample is presented in Figure 7(e). The sensitivity to the in-plane and cross-plane thermal conductivities of the GaN thin film are considerably lower when the spot sizes are 20 $\mu$m compared to the 10 $\mu$m spot sizes. The cross-plane thermal conductivity of the GaN thin film is measured by TDTR. At room temperature, the in-plane and cross-plane thermal conductivity of the GaN thin film can be considered to the same.\cite{lindsay2012thermal} The SSTR-measured thermal conductivity of the GaN substrate is 194 $\pm$ 27 W m$^{-1}$ K$^{-1}$ when the spot sizes are 10 $\mu$m. Using spot sizes of 20 $\mu$m, the thermal conductivity of the GaN substrate is measured with a lower uncertainty to be 185 $\pm$ 16 W m$^{-1}$ K$^{-1}$. 

The thermal conductivity of the sapphire substrate is measured by SSTR in the $\sim$2 $\mu$m AlN thin film on sapphire sample. The sensitivity calculation for this sample is shown in Figure 7(f). SSTR measurement of the sapphire substrate thermal conductivity is most sensitive to the in-plane thermal conductivity of the AlN thin film. The cross-plane thermal conductivity of this AlN thin film is measured by TDTR. As the anisotropy in the AlN thermal conductivity of is very small at room temperature,\cite{lindsay2013ab} the in-plane and cross-plane thermal conductivities of the 2 $\mu$m AlN thin film can be assumed to be the same. Using SSTR, the thermal conductivity of the sapphire substrate is measured to be 35.1 $\pm$ 5.9 W m$^{-1}$ K$^{-1}$ with 1/e$^{2}$ pump and probe radii of 10 $\mu$m. Similar to the other two samples, with 20 $\mu$m spot sizes, the sapphire thermal conductivity can be determined with a lower uncertainty, 34.5 $\pm$ 4.2 W m$^{-1}$ K$^{-1}$.

In Table 2, we present the measured substrate thermal conductivities for the two spot sizes. The uncertainty of the measured values incorporate the uncertainty associated with the $\gamma$ value (sapphire reference), Al transducer and thin film thermal conductivity, thin film thickness, and the thermal boundary conductances. The values of these parameters are tabulated in Table 1. As shown in Table 2, the SSTR-measured substrate thermal conductivities are in excellent agreement with literature. This proves the ability of SSTR to accurately measure the thermal conductivities of buried substrates that are typically inaccessible by TDTR and FDTR.

\subsection{Experimental measurement of the thermal conductivity of buried films by SSTR}

\begin{figure*}[hbt!]
\centering
\includegraphics[scale=0.4]{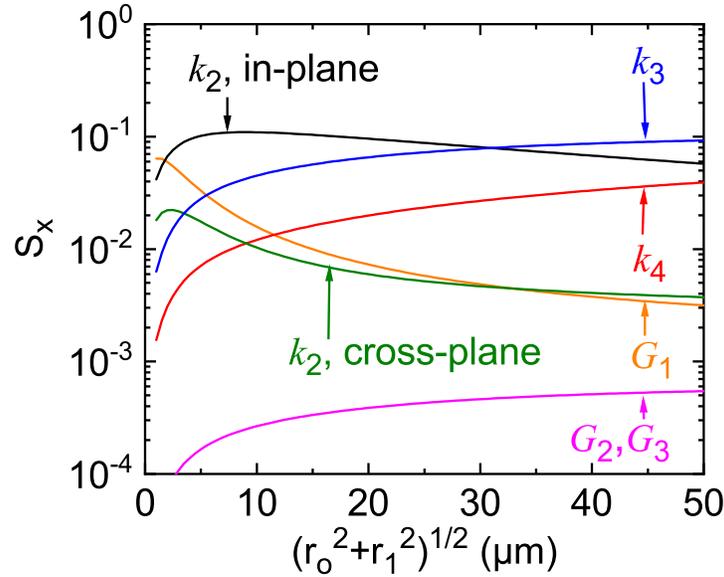} 
\caption{Sensitivity calculation as a function of effective radius, $\sqrt{r{_o^2}+r{_1^2}}$ for the 4-layer sample: 85 nm Al transducer/2.5 $\mu$m Si film/1 $\mu$m SiO$_2$ layer/Si substrate.
}
\label{fig:6}
\end{figure*}

We now discuss the required criteria for SSTR to measure the thermal conductivity of a buried film in a 4-layer system: metal transducer/thin film/buried film/substrate. Measurement of such a buried film is possible when the thermal resistance of this layer is much greater those of the top thin film and substrate. This stems from the fact that for SSTR to measure the thermal conductivity of any layer in a multilayered material system, a significant steady-state temperature gradient must exist in that layer, either in cross-plane or in-plane direction. As the top thin film is in contact with the metal transducer, the temperature gradient of this layer is often large unless the film thickness is very low. On the other hand, since the substrate is a semi-infinite medium, a measurable temperature gradient exists in the substrate when large pump and probe radii are used. For a buried film, however, unless the thermal resistance is large, the resulting temperature gradient is relatively small compared to those of the thin film and substrate. Therefore, although SSTR probes through the buried film and is influenced by the thermal properties of this layer, the degree of such influence is also relatively small. As a result, SSTR can not isolate the thermal conductivity of a buried film with low thermal resistance. 

In addition, large pump and probe radii (> 10 $\mu$m) are needed for buried film measurements. When the thermal resistance of the buried layer is much higher than those of thin film and substrate, bulk of the heat flows along the in-plane direction of the top thin film. For a sufficient thermal gradient to exist in the buried film, large spot sizes are required. 

To experimentally show this, we have selected a sample that fits this criteria: 85 nm Al transducer/2.5 $\mu$m Si film/1 $\mu$m SiO$_2$ layer/Si substrate. The sensitivity calculation for this sample is shown in Figure 8. As exhibited here, SSTR can measure the thermal conductivity of buried SiO$_2$ layer when large spot sizes are used. However, such measurements are also sensitive to the in-plane thermal conductivity of the top Si film. TDTR is used to measure the cross-plane thermal conductivity of top Si film as shown in Table 1. The in-plane and cross-plane thermal conductivity of the 2.5 $\mu$m Si film can be considered to the same.\cite{dong2015ballistic} 
Using 1/e$^2$ pump and probe radii of $\sim$20 $\mu$m, we measure the buried SiO$_2$ film thermal conductivity to be 1.34 $\pm$ 0.26 W m$^{-1}$ K$^{-1}$. This value is in agreement with literature,\cite{cahill1990thermal,giri2020ultralow} showing the capability of SSTR to measure the thermal conductivity of sub-surface buried layers. \\

\textbf{\Large Conclusion}

We experimentally and numerically investigate the influences of multilayer material systems, thin metal film transducers, and thermal boundary conductances on the TPD of SSTR technique. The traditional TPD definition of 1/e temperature drop distance from the maximum surface temperature does not represent the absolute upper limit of SSTR probing depth. Thus, when estimating whether the thermal conductivity of a buried substrate is measurable within acceptable limits of uncertainty, sensitivity calculations provide the best means. The low modulation freqency of SSTR enables it to measure the thermal conductivity of buried substrates typically inaccessible by TDTR and FDTR, demonstrated by presenting experimental data on three control samples. In addition, SSTR has the capability to isolate the thermal properties of a buried film as long as the thermal properties of this layer is much higher than those of the top thin film and substrate. This work marks an advancement in experimental metrology by establishing SSTR as a robust technique for thermal characterizations of subsurface buried substrates. 

\medskip
\textbf{Acknowledgements} \par 
The authors would like to acknowledge the financial support from U.S. Office of Naval Research under a MURI program (Grant no. N00014-18-1- 5332429). Z. C. Leseman acknowledges support provided by the Deanship of Scientific Research at King Fahd University of Petroleum $\&$ Minerals for funding this work through project SR191001.

\medskip

%
\bibliographystyle{achemso}
\bibliography{References}




\end{document}